\documentclass[journal]{IEEEtran}

\usepackage[utf8]{inputenc} 
\usepackage{url}
\usepackage{ifthen}
\usepackage{multirow}
\usepackage{cite}
\usepackage[cmex10]{amsmath} 
\usepackage{multicol}
\usepackage[T1]{fontenc}
\usepackage{setspace,caption}
\usepackage[linesnumbered,ruled,vlined]{algorithm2e}
\usepackage{algpseudocode}
\usepackage{amsfonts, amsthm, amssymb, wasysym,bm, bbm,graphicx,mathtools} 
\DeclareMathOperator*{\dprime}{\prime \prime}
\DeclareMathOperator*{\tprime}{\prime \prime \prime}
\usepackage{epstopdf,multicol} 
\usepackage[font=small,labelfont=bf]{caption}
\usepackage{pdfpages, float}
\usepackage{mathrsfs}
\usepackage{graphicx}
\usepackage[aboveskip=-1ex]{subcaption}

\newtheorem{lem}{Lemma}

\newtheorem{rem}{Remark}
\newtheorem{assumption}{Assumption}

\newtheorem{definition}{Definition}

\usepackage[linesnumbered,ruled,vlined]{algorithm2e}
\usepackage{algpseudocode}

\theoremstyle{definition}
\newtheorem{exmp}{Example}

\newtheorem{prop}{Proposition}
\usepackage{subcaption}

\DeclarePairedDelimiter{\floor}{\lfloor}{\rfloor}

\definecolor{blu}{RGB}{65,105,225}
\definecolor{purp}{RGB}{128,0,128}
\definecolor{rd}{RGB}{255,69,0}

\date{}
\begin{document}

\title{
	Nested Array-Based \\
	Spatially Coupled LDPC Codes 
	\thanks{This material is based on work supported by the National Science Foundation under Grant Nos. ECCS-1710920, ECCS-1711056, OIA-1757207, and HRD-1914635. This paper has been presented in part at the 2019 IEEE Information Theory Workshop, Visby, Sweden \cite{HMK19}.}}

\author{
	\begin{small}
	\IEEEauthorblockN{Salman Habib$^\dagger$, David G. M. Mitchell$^\ast$, and J{\"o}rg Kliewer$^\dagger$\\}
	\IEEEauthorblockA{$^\dagger$Helen and John C.~Hartmann Dept. of Electrical and Computer Engineering, \\
		New Jersey Institute of Technology, Newark, NJ 07102\\
		$^\ast$Klipsch School of Electrical and Computer Engineering, New Mexico State University, Las Cruces, NM 88003\\
	}
	\end{small}
}

\maketitle

%
\begin{abstract} 
	Linear nested codes, where two or more sub-codes are nested in a global code, have been proposed as candidates for reliable multi-terminal communication. In this paper, we consider nested array-based spatially coupled low-density parity-check (SC-LDPC) codes and propose a line-counting based optimization scheme for minimizing the number of dominant absorbing sets in order to improve its performance in the high signal-to-noise ratio regime. {Since the parity-check matrices of different nested sub-codes partially overlap, the optimization of one nested sub-code imposes constraints on the optimization of the other sub-codes. To tackle these constraints, a multi-step optimization process is applied first to one of the nested codes, then sequential optimization of the remaining nested codes is carried out based on the constraints imposed by the previously optimized sub-codes.} Results show that the order of optimization has a significant impact on the number of dominant absorbing sets {in the Tanner graph of the code}, resulting in a trade-off between the performance of a nested code structure and its optimization sequence: the code which is optimized without constraints has fewer harmful structures than the code which is optimized with constraints. We also show that for certain code parameters, dominant absorbing sets in the Tanner graphs of all nested codes are completely removed using our proposed optimization strategy.
\end{abstract}

\begin{small}
{\bf Keywords:} LDPC codes, belief propagation, nested codes, spatially coupled codes, absorbing sets, optimization. 
\end{small}


\section{Introduction}
\label{Intro}

Linear nested codes are error correcting codes where multiple information words are encoded separately and then are algebraically superimposed at the physical layer prior to transmission \cite{XFKC07}, \cite{KDH07}. Nested codes are commonly used for error correction in wireless multi-terminal networks. In nested coding, each information word is associated with a codeword which belongs to a different code, also called a \emph{sub-code}. Suppose, for example, there are two information words $\mathbf{u}_1$ and $\mathbf{u}_2$ of length $k_1$ and $k_2$, respectively. A codeword corresponding to each information word, $\mathbf{x}_i\in \mathbb{F}_2^n$, $i\in \{1,2\}$, is generated via a generator matrix $\mathbf{G}_i$, or parity-check matrix $\mathbf{H}_i$, respectively. Here, $\mathbf{x}_i$ belongs to a different sub-code $\mathscr{C}_i$, and the transmitted codeword $\mathbf{x}$ is an element of the \emph{global} code $\mathscr{C}$. For example, the generator matrix $\mathbf{G}$ of the global code is obtained by stacking the individual $\mathbf{G}_1$ and $\mathbf{G}_2$ matrices vertically as $\left[\begin{smallmatrix} \mathbf{G}_1 \\ \mathbf{G}_2 \\ \end{smallmatrix}\right]$.


\vspace{0.1cm}
The construction of nested regular and irregular low-density parity-check block codes (LDPC-BCs) have been considered in \cite{KK10} and \cite{GM18}. Irregular LDPC-BCs {can be optimized to} outperform regular LDPC-BCs in the waterfall region {under belief propagation (BP) decoding}; however, they {generally} suffer from an early onset of an error-floor, a flattening of the bit error rate (BER) performance curve in the high signal-to-noise ratio (SNR) region, and are not implementation friendly \cite{CDFKMS14}. In comparison, regular LDPC-BCs {generally} have better error-floor performance {as a result of their minimum distance and graph properties \cite{Gal62,R03}} and lower decoder complexity; however, their performance deteriorates with increasing graph density {under BP decoding} \cite{RU08}, making them undesirable for the construction of nested codes that require a higher graph density than the non-nested ones. 

Spatially coupled LDPC (SC-LDPC) codes \cite{CDFKMS14,YFZLMC20}, on the other hand, are known to approach the capacity of binary input memoryless channels under {BP decoding} as the graph density increases \cite{KRU13}. Hence, regular SC-LDPC codes are good potential candidates for nested code constructions. SC-LDPC codes are obtained by coupling, or connecting, multiple Tanner graphs corresponding to an underlying LDPC-BC. One potential obstacle of graph-based codes is that Tanner graphs contain small sub-structures called absorbing sets (ABSs) \cite{DAnantharam10} which are known to cause the BP decoder to fail. These failures are responsible for the error-floor phenomenon. However, since spatial coupling is able to {reduce or eliminate many of} these harmful ABSs \cite{MDC14}, SC-LDPC codes have superior error-floor performance when compared to their BC {counterparts}. 

{The optimization of SC-LDPC codes for magnetic recording channels is considered in \cite{HWD20}, which generalizes \cite{EHD18} for ABSs of type $(4,4(\gamma-2))$, where $\gamma$ is the column weight. Moreover, \cite{BCCB18} analyzes the conditions to avoid cycles of lengths $6$ and $8$ in SC-LDPC codes. An array-based (AB) SC-LDPC code is constructed from the Tanner graph of an AB-LDPC-BC by applying an edge-spreading technique \cite{ARKD15},\cite{MR17},\cite{EHD18}. AB-SC-LDPC codes possess a regular quasi-cyclic (QC) structure that makes them more attractive for hardware implementation because different regions on the Tanner graph of AB-LDPC codes can be decoded in parallel, which improves the decoding throughput and lowers the decoding latency. Moreover, their structure also guarantees a certain minimum distance \cite{R15} and a girth of $6$ (they are free of 4-cycles). These features, in turn, guarantee the non-existence of certain harmful ABSs \cite{DAnantharam10}.} Due to these advantages, we investigate the construction of nested AB-SC-LDPC codes, which, to the best of our knowledge, has not been addressed in the open literature so far.  

Our objective for finite length nested code design is to ensure that each nested sub-code and the global code have a small number of dominant ABSs in the Tanner graph when compared to the underlying LDPC-BCs. In this paper, we propose an adapted line-counting (ALC) technique to optimize the design of nested AB-SC-LDPC codes. In contrast to the line counting approach of \cite{ASCJ17}, the presented ALC technique allows the enumeration of 6-cycles in arbitrary column weight-$3$ sub-matrices (of nested parity-check matrices) that form the dominant ABSs in polynomial time, facilitating a tractable nested code optimization. Since the parity-check matrices of different nested sub-codes partially overlap, the null-spaces for these sub-codes intersect. Consequently, an optimization of  one nested sub-code affects other nested sub-codes, and thus imposes constraints on the optimization. Since multiple design constraints must now be jointly satisfied, the construction of nested codes for the multi-terminal setting is more challenging than for the point-to-point case. We demonstrate that, by using ALC, it is possible to minimize/eliminate dominant ABSs in certain nested AB-SC-LDPC codes, irrespective of the row and column weight of the overall parity-check matrix containing all the nested matrices. We also show that, for certain code parameters, dominant absorbing sets in the Tanner graphs of all nested codes are completely removed using our proposed optimization strategy. Simulation results are provided that confirm the improved nested code performance promised by the approach.



\section{Preliminaries}
\label{sec:prelims}
\subsection{Protograph and Array-Based LDPC Codes}
\label{sec:alg_lft}

An LDPC-BC with parity-check matrix $\mathbf{H}\in \mathbb{F}_2^{l \times n}$, where $l\geq n-k$ and $k$ is the length of a message block, is often designed based on a protograph \cite{Thorpe03,FCCLLH19}, which is a small Tanner graph consisting of $p$ VNs and $\gamma$ CNs, $p\geq \gamma$, with a \emph{design rate} given by $R=1-\frac{\gamma}{p}$. Let $\mathbf{B}=[B_{i,j}]_{\gamma\times p}$ represent the \emph{base} matrix corresponding to the Tanner graph of protograph $G_{\mathbf{B}}$, where $B_{i,j}$ denotes the number of edges connecting CN $i$ to VN $j$. By applying a graph lifting procedure with lifting factor $p$, that is by replacing each non-zero entry of $\mathbf{B}$ with a sum of $B_{i,j}$ non-overlapping permutation matrices of size $p\times p$, and each zero entry with an all-zero matrix of size $p\times p$, we construct an LDPC  matrix $\mathbf{H}\in \mathbb{F}_2^{\gamma p \times p^2}$. The resulting rate is $R\geq 1-\frac{\gamma}{p}$, with equality if and only if the parity-check matrix is full-rank. In the case of an AB-LDPC-BC, $\mathbf{B}$ is a $\gamma\times p$ all-ones matrix, where $p$ is prime \cite{Fan00}. {The parity-check matrix $\mathbf{H}(\gamma,p)$ of an AB-LDPC-BC parity-check matrix consists of circulant matrices, where the entries of a circulant matrix $\mathbf{\sigma}^z$ is obtained by circularly left-shifting the non-zero entries of the identity matrix $\mathbf{I}$ by $z \text{ mod } p$ \cite{DAnantharam10,AREKD16}. Note that $\gamma$ is also the column weight of $\mathbf{H}(\gamma,p)$.}

%
%


Let $q\in\{0,1,\ldots,\gamma-1\}$ and $s\in\{0,1,\ldots,p-1\}$ denote the row group number, and the row number within a particular row group of $\mathbf{H}(\gamma,p)$, respectively. For example, by $q=0$ we refer to the row group of $\mathbf{I}$ matrices in $\mathbf{H}(\gamma,p)$. Also, let $j\in\{0,1,\ldots,p-1\}$ and $k\in\{0,1,\ldots,p-1\}$ represent the column group number and the column number inside a particular column group of an AB-LDPC matrix $\mathbf{H}(\gamma,p)$. Therefore, each row (resp., column) of an AB matrix is given by $r=qp+s$ (resp., $c=jp+k$). In this way, the location of an entry of an AB matrix $(r,c)$ may be written as $(q,s;j,k)$. Note that when referring to multiple row groups, we use subscripts $q_0$, $q_1,\ldots,$ and so on where $q_i\in \{0,1,\ldots,\gamma-1\}$, with similar subscript use for multiple row numbers, column groups, and column numbers. 

\vspace{-0.25cm}
\subsection{Array-Based SC-LDPC Codes}
	AB-SC-LDPC codes can be constructed from AB-LDPC-BCs by coupling $L$ copies of the Tanner graph $G_{\mathbf{H}}$ via edge-spreading \cite{MLC15}. In terms of matrices, edge-spreading is equivalent to splitting $\mathbf{H}(\gamma,p)$ into a sum of $m+1$ component block matrices of the same dimension as $\mathbf{H}(\gamma,p)$, such that $\mathbf{H}(\gamma,p)=\mathbf{H}_0+\mathbf{H}_1+\cdots+\mathbf{H}_m$, where $m$ denotes the \emph{memory} of the code. The construction involves a spreading matrix $\mathbf{B}_m\in \mathbb{F}_{m+1}^{\gamma \times p}$, where an entry $g\in \{0,1,\ldots,m\}$ in position $(i,j)$ of this matrix indicates that the $p\times p$ circulant block in row group $i$, column group $j$, of $\mathbf{H}(\gamma,p)$ is copied to its corresponding position in $\mathbf{H}_g$ \cite{MR17}. The resulting time-invariant parity-check matrix of a terminated SC-LDPC code is denoted $\mathbf{H}(\gamma, p, L)\in \mathbb{F}_2^{\gamma p(L+m)\, \times\, Lp^2}$ and is given as
	
	
\begin{align}
\label{eq:sc_mat}
\mathbf{H}(\gamma,p,L)=
\begin{bmatrix}
\mathbf{H}_0 & & \\
\mathbf{H}_1 & \ddots &  \\
\vdots & \ddots & \mathbf{H}_0 \\
\mathbf{H}_m &  & \mathbf{H}_1 \\
& \ddots & \vdots \\
& & \mathbf{H}_m
\end{bmatrix},
\end{align}

\vspace{-0.25cm}
\noindent where $L>m+1$ is the number of \emph{column blocks} (each containing $p$ column groups), called the \emph{coupling length}, and $\nu=(m+1)p^2$ is the \emph{constraint length} of $\mathbf{H}(\gamma,p,L)$. We can also label the row and column blocks $y\in \{0,1,\ldots,L+m-1\}$ and $v\in \{0,1,\ldots,L-1\}$ in a similar way to AB-LDPC-BCs, where an individual entry $(r,c)$ in {an AB-SC-LDPC} code may then be written as  $(y,q,s; v,j,k)$. 


\vspace{-0.1cm}
\subsection{Nested Codes}
\vspace{-0.1cm}
A nested code consists of a group of $M$ sub-codes $\mathscr{C}_i$, $i=1,2,\ldots,M$, $M\geq 2$, nested in a global code $\mathscr{C}$ of rate $k/n$ with the property $\mathscr{C}_i\subset \mathscr{C}$, $\forall i$. Nested codes are used to jointly encode $M$ different information vectors $\mathbf{u}_i\in \mathbb{F}_2^{k_i}$, $k_i<k$, to generate an overall codeword $\mathbf{x}\in \mathbb{F}_2^n$, which is a linear combination of all the codewords $\mathbf{x}_i$ obtained from each of the sub-codes. The process of obtaining $\mathbf{x}$ via nested coding is expressed as 

\vspace{-0.75cm}
\begin{align*}
	\mathbf{x}^T&=[\mathbf{u}_1^T,\mathbf{u}_2^T,\ldots,\mathbf{u}_M^T]
	\left[
		\begin{array}{c} 
			\mathbf{G}_1 \\
			\mathbf{G}_2 \\
			\vdots \\
			\mathbf{G}_M \\
		\end{array} 
	\right]=[\mathbf{u}_1^T,\mathbf{u}_2^T,\ldots,\mathbf{u}_M^T]\mathbf{G} \\
	&=\mathbf{u}_1^T\mathbf{G}_1\oplus \mathbf{u}_2^T\mathbf{G}_2\oplus \cdots \oplus \mathbf{u}_M^T\mathbf{G}_M =\mathbf{x}_1^T\oplus \mathbf{x}_2^T\oplus \cdots\oplus \mathbf{x}_M^T,
\end{align*}

\noindent where $\oplus$ denotes the bitwise XOR operation and $\mathbf{G}_i\in \mathbb{F}_2^{k_i\times n}$, $\mathbf{G}\in \mathbb{F}_2^{k\times n}$ are the generator matrices of sub-code $\mathscr{C}_i$ and global code $\mathscr{C}$, respectively, with $k=\sum_{i=1}^M{k_i}$. Note that each codeword $\mathbf{x}_i$ is encoded at a rate $k_i/n$, whereas the overall codeword $\mathbf{x}$ is encoded at a rate $k/n$. 

{In comparison, nested dual codes of the above mentioned codes are defined using parity check matrices $\mathbf{H}_i\in \mathbb{F}_2^{l_i \times n}$, $l_i\geq n-k_i$, and $\mathbf{H}\in \mathbb{F}_2^{l \times n}$, $l\geq n-k$, corresponding to the nested codes $\mathscr{C}_i$ and $\mathscr{C}$, respectively, which form the null spaces of matrices $\mathbf{H}_i$ and $\mathbf{H}$, respectively.} Note that the matrices $\mathbf{H}_i$, $\forall i$, and $\mathbf{H}$ are considered as (potentially overlapping) sub-matrices of a larger $\mathbf{\hat{\mathbf{H}}}\in \mathbb{F}_2^{b\times n}$ matrix, where $l_i< b\leq n \text{ },\forall i$, and $l<b\leq n$, respectively. The construction details of $\mathbf{H}_i$ and $\mathbf{H}$ from $\mathbf{\hat{\mathbf{H}}}$ are discussed in \cite{KK10} for LDPC-BCs. 

{In the remainder of the paper, we consider $\mathbf{H}_i$ and $\mathbf{H}$ to be the parity-check matrices of nested regular LDPC codes.} We refer to the column weight $\omega_i$ (resp., $\omega$) of sub-code $\mathscr{C}_i$ (resp., global code $\mathscr{C}$) as the column weight of its corresponding (regular) parity-check matrix $\mathbf{H}_i$ (resp., $\mathbf{H}$). This is the construction used in the rest of the paper.


\subsection{Absorbing Sets}
\vspace{-0.1cm}

In $G_{\mathbf{H}}$, suppose $X\subset V$ and let $\mathcal{N}(X)$ be the set of all neighbors of $X$. Let $O(X)$ be the set of neighbors of $X$ with odd degree in the subgraph induced by $X\cup \mathcal{N}(X)$. 

\begin{definition}[\hspace{-0.1cm} \cite{DAnantharam10}]
	{For $a>1$, $b\geq 0$,} an $(a,b)$ ABS $X$ is a set of VNs with $|X|=a$, $|O(X)|=b$, and the property that each VN in $X$ has strictly fewer neighbors in $O(X)$ than in $C\setminus O(X)$. An $(a,b)$ ABS is a fully ABS if, additionally, all nodes in $V\setminus X$ have strictly {fewer neighbors in $O(X)$ than in $C\setminus O(X)$.} A minimal $(a,b)$ ABS refers to an ABS which has the smallest possible existing value for $a$ in a given LDPC Tanner graph, and where $b$ is the smallest possible value for the given $a$.
\end{definition}

\begin{definition}
A block $k$-cycle in an AB-LDPC code is a collection of $p$ cycles of length $k$, where each cycle in the collection spans the same row and column groups of the parity-check matrix.
\end{definition}

\begin{rem}
	\label{rem:6cyc_33ABS}
	{A $(3,3)$ ABS is the minimal ABS that can exist in a column weight-$3$ AB-LDPC-BC code \cite{DAnantharam10}}. For $\gamma=4$, a $(4,4)$ ABS is the minimal ABS in an AB-LDPC-BC for $p=5,7$; whereas, a $(5,4)$ ABS is the minimal ABS in an AB-LDPC-BC for the case {$p=11,19$}; finally, $(6,4)$ ABSs exist in an AB-LDPC-BC for $p>5$, and it is the minimal ABS for {all} $p>19$ \cite{DAnantharam10}. For $\gamma=5$, the {minimal ABS in AB-LDPC-BCs is of size $(4,8)$, and $(5,9)$ and $(6,8)$ ABSs are also dominant} \cite{WDW13}.\hspace{-0.05cm}\footnote{By ``dominant'', we mean those ABS(s) which are empirically observed to cause the majority of failures in the high SNR regime.}  
\end{rem}


\noindent {From the structure of dominant ABSs discussed in \cite{DAnantharam10}, \cite{WDW13}, we note the following.}

\begin{rem}
	\label{rem:64abs}
	{For $\gamma=3$, a $(3,3)$ ABS in an AB-LDPC code corresponds to a 6-cycle and a $(4,2)$ ABS consists of two $(3,3)$ ABSs. Moreover, for} $\gamma=4$ (resp., $\gamma=5$), the $(4,4)$, $(5,4)$ and $(6,4)$ (resp., $(4,8)$, $(5,9)$ and $(6,8)$) ABSs in an AB-LDPC code all contain at least one 6-cycle. {Consequently, by minimizing the number of 6-cycles in AB-SC-LDPC codes, we also minimize the related ABSs discussed in Remark \ref{rem:6cyc_33ABS}.}
\end{rem}


\subsection{Line-counting}     
\label{sec:LC}

\begin{figure}[h]
\centering
\includegraphics[scale=2.5]{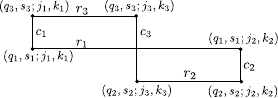}
\vspace{-0.1cm}
\caption{The structure of a 6-cycle in an AB-LDPC matrix.}
\vspace{-0.5cm}
\label{fig:6-cycle}
\end{figure}

A 6-cycle must span three distinct row and column groups of an AB matrix \cite{ASCJ17,BCCB18}. As shown in Fig. \ref{fig:6-cycle}, suppose that the columns of a 6-cycle have indices $c_1,c_2,c_3$ and that they exist in distinct column groups $j_1,j_2,j_3$, respectively. Similarly, suppose that the rows of a 6-cycle have indices $r_1,r_2,r_3$ and they exist in distinct row groups $q_1,q_2,q_3$, respectively. In \cite{AREKD16}, $(3,3)$ ABSs are enumerated by computing the area of a 2D polytope lying on the $(j,j')$ plane, $j\neq j'$, where a data point (coordinate) in this polytope corresponds to a $(3,3)$ ABS (and hence a 6-cycle).\hspace{-0.025cm}\footnote{{We note that line counting is performed on column weight-$3$ parity-check matrices throughout the paper, consequently there is an equivalence between 6-cycle and a $(3,3)$ ABS and we refer to them interchangeably. In general, a 6-cycle corresponds to a $(3,3(\gamma-2))$ trapping set in a column weight $\gamma$ parity-check matrix \cite{EHD18}.}} This technique, however, has two drawbacks: first, it is only applicable to AB-SC-LDPC codes constructed via the \emph{cutting-vector} scheme \cite{MDC14}, and secondly, it only works for column weight-$3$ AB-LDPC codes. To allow more general codes, such as the ones obtained via general edge-spreadings \cite{MR17}, a $(3,3)$ ABS enumeration technique, namely line-counting, was proposed in \cite{ASCJ17} which enumerates $(3,3)$ ABSs by counting 6-cycles on the $(c_1,c_2)$ plane.


In this paper, we propose a variant of line-counting, called adapted line-counting (ALC), in conjunction with an optimization algorithm (more details in Section \ref{sec:Opt}) to recursively optimize the entries of a $\mathbf{B}_m$ spreading matrix in order to ensure that the resulting $\mathbf{H}(\gamma,p,L)$ matrix contains as few harmful ABSs as possible. In order to facilitate presentation of our ALC scheme later in Section \ref{sec:nested}, the line-counting method of \cite{ASCJ17} is briefly reviewed. We refer to a \emph{matrix region} $\mathcal{R}$ that consists of at least six (not necessarily contiguous) circulant matrices spread across three row groups and at most $p$ column groups, with one row group being a row group of $\mathbf{I}$ matrices only. A row block from a $\mathbf{H}(\gamma=3,p,L)$ matrix in the case of AB-SC-LDPC codes, contains a row group of $\mathbf{I}$ matrices and two row groups consisting of $\sigma^{fz \text{ mod }p}$ matrices for $0\leq z\leq p-1$ and $f=1,2$. W.l.o.g., for any 6-cycle in $\mathcal{R}$, we have the following \cite{ASCJ17} (row and column block indices in the case of AB-SC-LDPC codes are fixed and dropped for clarity):


\vspace{-0.1cm}
\begin{itemize}
	\item $c_2>c_1$ and $w_1 p\leq c_1 <w_2 p$, $w_3 p\leq c_2<w_4 p$, where $w_1,w_2,w_3,w_4$ are integers satisfying {$0\leq w_1\leq p-2$}, $1\leq w_2\leq p-1$, {$w_1+1\leq w_3\leq p-1$}, and $w_2+1\leq w_4\leq p$. If the 6-cycle row $r_1$, incident to columns $c_1$ and $c_2$ (see Fig. \ref{fig:6-cycle}), exists in a row group of $\mathbf{I}$ matrices, we obtain $c_2-c_1=np$, where $n=\{1,2,\ldots,w_4-w_1-1\}$;
	\vspace{-0cm}
	\item $\alpha p\leq c_3 < \beta p$, where $\alpha$ and $\beta$ are integers satisfying $0\leq\alpha\leq p-1$, $1\leq\beta\leq p$, and $\alpha<\beta$. 
\end{itemize} 


The circulant matrix $\sigma^{z}$, $z\in\{0,1,\ldots,p-1\}$, has its non-zero elements located at $(s,k)$, where $s = z+k \text{ mod }p$. W.l.o.g., let $q_2$ (resp., $q_3$) represent the index of the row group containing all the $\sigma^{2z \text{ mod }p}$ (resp., $\sigma^{z}$) circulant matrices ($q_2=2$ and $q_3=1$ in an AB-LDPC-BC). The edges (ones in the parity-check matrix) corresponding to $(r_2,c_3)$ and $(r_2,c_2)$ are in rows $s_2= 2j_3+k_3 \text{ mod }p$ and $s_2= 2j_2+k_2 \text{ mod }p$ within their row group $r_2$. Since these rows are identical, $2j_2+k_2 = 2j_3+k_3 \text{ mod }p$, which are rearranged as 

\vspace{-0.25cm}
\begin{equation}
k_3-k_2= 2j_2-2j_3 \text{ mod }p.\label{eq:1}
\end{equation} 

\vspace{-0.25cm}
\noindent Now, considering $(r_3,c_1)$, we note that, $s_3 = j_1+k_1 \text{ mod }p$, and $s_3= j_3+k_3 \text{ mod }p$. Since these rows are also identical, we get $j_1+k_2= j_3+k_3 \text{ mod }p$, which are rearranged as 
 
\vspace{-0.25cm}
\begin{equation}
j_1-j_3= k_3-k_2\text{ mod }p.\label{eq:2}
\end{equation} 

\vspace{-0.2cm}
\noindent Substituting (\ref{eq:1}) in (\ref{eq:2}), and after rearranging, we obtain

\vspace{-0.25cm}
\begin{equation}
j_3 = 2j_2-j_1 \text{ mod }p.\label{eq:3}
\end{equation} 

\vspace{-0.2cm}
\noindent Now, if the roles of $q_2$ and $q_3$ are interchanged, we obtain (via a similar procedure) 

\vspace{-0.25cm}
\begin{equation}
j_3= p+2j_1-j_2\text{ mod }p.\label{eq:4}
\end{equation} 

\vspace{-0.25cm}
\noindent Finally, by invoking the  inequality $\alpha \leq j_{3} \leq \beta-1$ and by taking into account the corresponding 6-cycle column value $c_3$ as a function of $c_1$ and $c_2$, we obtain the range of $c_3$ in $\mathcal{R}$ as \cite{ASCJ17}

\vspace{-0.725cm}
\begin{subequations}
\label{eq:ineq1}
\begin{align}
\label{eq:ineq1a}
\frac{\alpha p}{2}\leq c_{2}-\frac{1}{2}c_{1}<\frac{\beta p}{2},
\hspace{0.3cm} \frac{p^{2}+\alpha p}{2}\leq c_{2}-\frac{1}{2}c_{1}<\frac{p^{2}+\beta p}{2}, \\
\label{eq:ineq1b}
p^{2}-\beta p<c_{2}-2c_{1}\leq p^{2}-\alpha p, \text{ or}
\hspace{0.3cm} -\beta p<c_{2}-2c_{1}\leq-\alpha p,
\end{align}
\end{subequations}

\vspace{-0.25cm}
\noindent where the inequalities in (\ref{eq:ineq1a}) are obtained from (\ref{eq:3}), and the ones in (\ref{eq:ineq1b}) are obtained from (\ref{eq:4}).

\begin{figure*}[t]
\centering
\includegraphics[scale=0.475]{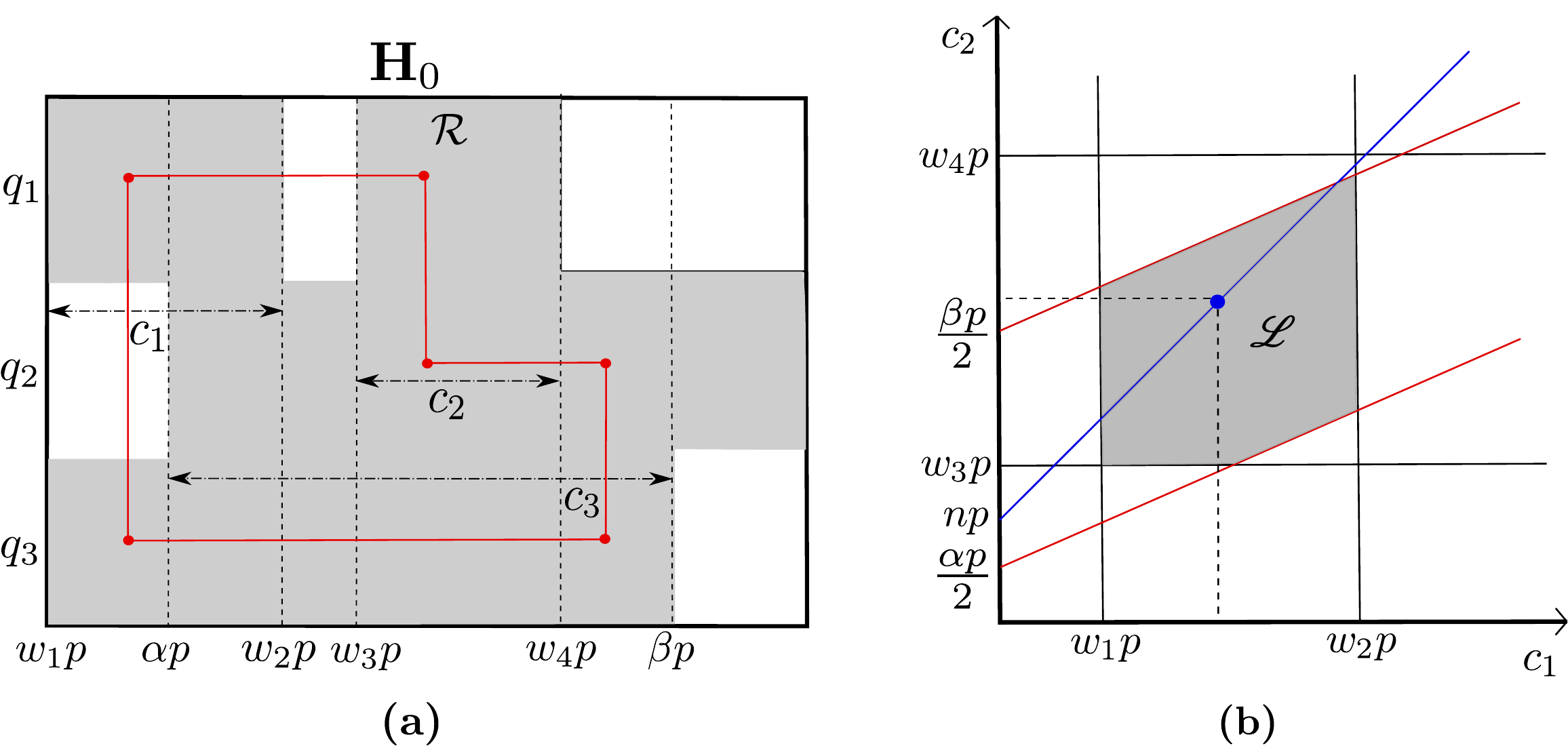}
\caption{An example of a {configuration of a} 6-cycle {corresponding to columns $c_1$, $c_2$, and $c_3$} in region $\mathcal{R}$ is shown in part (a), where the white region indicates all-zero circulant matrices {and the horizontal arrows represent the range of values of the columns $c_1$, $c_2$, and $c_3$ of this 6-cycle configuration. Part (b) shows the corresponding boundaries for such a configuration with these ranges, where a blue dot on the line $c_2-c_1=np$ within the shaded region $\mathscr{L}$ corresponds to this 6-cycle.}}
\label{fig:lc}
\end{figure*} 

{Consider Fig. \ref{fig:lc} for an illustration. The grey area in Fig. \ref{fig:lc}(a) is an example of a region $\mathcal{R}$ in an $\mathbf{H}_0$ matrix containing a 6-cycle (in red) with column (resp., row group) indices $c_1$, $c_2$, and $c_3$ (resp., $q_1$, $q_2$, and $q_3$) of $\mathbf{H}_0$. The white region in Fig. \ref{fig:lc}(a) indicates all-zero circulant matrices, and the horizontal arrows represent the range of values of the columns  $c_1$, $c_2$, and $c_3$  of the 6-cycle. These ranges generate vertical, horizontal and diagonal boundaries, respectively, on the $(c_1,c_2)$ plane in Fig. \ref{fig:lc}(b), and the area enclosed within these boundaries is shown in grey.} In particular, the red diagonal boundaries are obtained from the first inequality in (\ref{eq:ineq1a}). A 6-cycle with columns $c_1$ and $c_2$ exists in $\mathcal{R}$ if a coordinate $(c_1,c_2)$ on the line $c_2-c_1=np$ lies within the grey region $\mathscr{L}$ on the $(c_1,c_2)$ plane. Consequently, the number of 6-cycles in $\mathcal{R}$ is determined by the number of integer points on the line $c_2-c_1=np$ within $\mathscr{L}$. 


{Note that the inequalities in (\ref{eq:ineq1a}) and (\ref{eq:ineq1b}) only hold for column weight-$3$ AB matrices where the circulant matrix shift factor is $f=1,2$. Consequently, the line-counting method of \cite{ASCJ17} is not applicable for arbitrary column weight-$3$ AB-LDPC-BC matrices (for {the} case $f>2$), which will be encountered during nested code optimization.} To allow {for} $f>2$, we introduce ALC in Section \ref{sec:ALC} which generalizes these inequalities. This will facilitate the nested code optimization considered in Section \ref{sec:Opt_Alc}.

\vspace{-0.15cm}
\section{Nested AB-SC-LDPC Codes}
\vspace{-0.15cm}
\label{sec:nested}

In this section, we discuss the construction procedure of nested AB-SC-LDPC codes from the corresponding nested AB-LDPC-BCs. For nested AB-SC-LDPC codes with $\omega_i=3,4$, or $5$, whose dominant ABSs are known to contain 6-cycles (see Remark \ref{rem:64abs}), ALC can be used directly as a tool in the construction. \hspace{-0.15cm}\footnote{If for $\omega_i>5$, the dominant ABSs in the code's Tanner do not contain 6-cycles, then ALC as stated here would not be applicable. Nonetheless, the method discussed in this section is still useful provided that a similar ABS enumeration scheme is employed for scenarios where the dominant ABSs in nested codes do not contain 6-cycles.}

\vspace{-0.15cm}
\subsection{General Construction}
\label{sec:general_cons}

\vspace{-0.2cm}
We form a nested AB-LDPC-BC sub-code $\mathscr{C}_i$, $i\in\{1,2,\ldots,M\}$, and a global code $\mathscr{C}$ from sub-matrices $\mathbf{H}_i(w_i,p)\in \mathbb{F}_{{2}}^{p\omega_i \times p^2}$ and $\mathbf{H}(3,p)$, respectively, where $\omega_i>3$. {Let $\Omega_i$ be the number of possible sub-codes $\mathscr{C}_i$ for column weight $\omega_i\geq 4$.} To simplify our analysis, we suppose that $\mathbf{H}_i(w_i,p)$ (resp., $\mathbf{H}(3,p)$) consists of row groups $(0,1,\ldots,\omega_i-2,\omega_i-1+h_i)$, $h_i\in \{0,1,\ldots,\Omega_i-1\}$, (resp., $(0,1,2)$) of $\mathbf{\hat{H}}$, but generalizations of this selection are possible. {It follows from the array structure \cite{Fan00} that there are $\Omega_i=\gamma-\omega_i+1$, $\gamma\geq 5$, possible sub-codes,} and hence $\gamma$ should be chosen sufficiently large to construct the desired $M$ sub-codes. Note that the design rate $R_i=1-\frac{\omega_i}{p}$ of $\mathscr{C}_i$ decreases with increasing $\omega_i$. As a result, constructing high rate nested codes with large $\omega_i$ also requires a relatively large $p$ compared to a nested code with smaller $\omega_i$. In turn, this also makes each sub-code optimization more computationally intensive as $M$ increases.

\begin{exmp}
\label{exmp:block}
We construct the $\mathbf{\hat{H}}(\gamma,p)$ AB-LDPC-BC matrix with $\gamma=5$, $p=5$, and $M=2$, where 

\vspace{-0.25cm}

\[\mathbf{\hat{H}}(5,5)=
		\left[
			\begin{array}{ccccc}
				\mathbf{I} & \mathbf{I}  & \mathbf{I}  & \mathbf{I}  & \mathbf{I} \\
				\mathbf{I} & \mathbf{\sigma} & \mathbf{\sigma^2} & \mathbf{\sigma^3} & \mathbf{\sigma^4} \\
				\mathbf{I} & \mathbf{\sigma^2} & \mathbf{\sigma^4} & \mathbf{\sigma} & \mathbf{\sigma^3} \\
				\color{blu}{\mathbf{I}} & \color{blu}{\mathbf{\sigma^3}} & \color{blu}{\mathbf{\sigma}} & \color{blu}{\mathbf{\sigma^4}} & \color{blu}{\mathbf{\sigma^2}} \\
				\color{red}{\mathbf{I}} & \color{red}{\mathbf{\sigma^4}} & \color{red}{\mathbf{\sigma^3}} & \color{red}{\mathbf{\sigma^2}} & \color{red}{\mathbf{\sigma}}  \\	
			\end{array}
\right].\] 
		
\noindent There are $\Omega_i=2$ possible sub-codes of column weight $\omega_i=4$ in our construction. This matrix consist of sub-matrices: $\mathbf{H}_1(4,5)$ (black and blue row groups $0$, $1$, $2$, and $3$, with $h_1=0$), $\mathbf{H}_2(4,5)$ (black and red row groups $0$, $1$, $2$, and $4$, with $h_2=1$), and $\mathbf{H}(3,5)$ (black row groups $0$, $1$, and $2$) with corresponding column weight $4$ nested sub-codes $\mathscr{C}_1$, $\mathscr{C}_2$, and a column weight-$3$ global code $\mathscr{C}$, respectively. Explicitly, the resulting nested sub-matrices are 


\vspace{-0.5cm}
\begin{align}
\begin{split}
     \mathbf{H}(3,5)&=
		\left[
			\begin{array}{ccccc}
				\mathbf{I} & \mathbf{I}  & \mathbf{I}  & \mathbf{I}  & \mathbf{I} \\
				\mathbf{I} & \mathbf{\sigma} & \mathbf{\sigma^2} & \mathbf{\sigma^3} & \mathbf{\sigma^4} \\
				\mathbf{I} & \mathbf{\sigma^2} & \mathbf{\sigma^4} & \mathbf{\sigma} & \mathbf{\sigma^3} \\
			\end{array}
		\right], \\
	\mathbf{H}_1(4,5)&=
			\left[
				\begin{array}{ccccc}
					\mathbf{I} & \mathbf{I}  & \mathbf{I}  & \mathbf{I}  & \mathbf{I} \\
					\mathbf{I} & \mathbf{\sigma} & \mathbf{\sigma^2} & \mathbf{\sigma^3} & \mathbf{\sigma^4} \\
					\mathbf{I} & \mathbf{\sigma^2} & \mathbf{\sigma^4} & \mathbf{\sigma} & \mathbf{\sigma^3} \\
					\color{blu}{\mathbf{I}} & \color{blu}{\mathbf{\sigma^3}} & \color{blu}{\mathbf{\sigma}} & \color{blu}{\mathbf{\sigma^4}} & \color{blu}{\mathbf{\sigma^2}} \\	
				\end{array}
			\right], \text{ and }\\
     \mathbf{H}_2(4,5)&=
		\left[
			\begin{array}{ccccc}
				\mathbf{I} & \mathbf{I}  & \mathbf{I}  & \mathbf{I}  & \mathbf{I}\\
				\mathbf{I} & \mathbf{\sigma} & \mathbf{\sigma^2} & \mathbf{\sigma^3} & \mathbf{\sigma^4} \\
				\mathbf{I} & \mathbf{\sigma^2} & \mathbf{\sigma^4} & \mathbf{\sigma} & \mathbf{\sigma^3} \\
				\color{red}{\mathbf{I}} & \color{red}{\mathbf{\sigma^4}} & \color{red}{\mathbf{\sigma^3}} & \color{red}{\mathbf{\sigma^2}} & \color{red}{\mathbf{\sigma}} \\	
			\end{array}
		\right].
\end{split}
\label{eq:3_mats}
\end{align} 
\vspace{-0.5cm}
\hfill\(\Box\)

\end{exmp}

\vspace{-0cm}
Nested AB-SC-LDPC matrices are constructed in a similar way, with parity-check matrices denoted as $\mathbf{H}_i(\omega_i,p,L)$ and $\mathbf{H}(3,p,L)$ obtained by edge-spreading $\mathbf{H}_i(\omega_i,p)$ and $\mathbf{H}(3,p)$ via spreading matrices $\mathbf{B}_{i,m}\in \mathbb{F}_{m+1}^{\omega_i \times p}$ and $\mathbf{B}_{m}\in \mathbb{F}_{m+1}^{3 \times p}$, respectively, where the row indices of $\mathbf{B}_{i,m}$ correspond to the row group indices of $\mathbf{H}_i(w_i,p)$. Since the codes are nested, the same relationship between $\mathbf{H}_i(w_i,p)$ and $\mathbf{H}(3,p)$ also holds for the edge-spreading matrices, \emph{i.e.}, $\mathbf{B}_{m}$ is a sub-matrix of $\mathbf{B}_{i,m}$. An example of creating nested AB-SC-LDPC codes from the nested AB-LDPC-BCs in Example \ref{exmp:block} is given below in Example \ref{exmp:sc1}.

\begin{exmp}
\label{exmp:sc1}

For memory $m=1$, we first generate an edge-spreading matrix

\vspace{-0.5cm}
\begin{align}
	\label{eq:B_1}
	\mathbf{B}_1=\left[
		\begin{array}{ccccc}
			1 & 0 & 0 & 0 & 1 \\
			1 & 1 & 1 & 0 & 0 \\
			0 & 0 & 1 & 1 & 0 \\	
		\end{array}
	\right]\in \mathbb{F}_2^{3\times p},
\end{align}

\vspace{-0.1cm}
\noindent which can be used to form global AB-SC-LDPC matrix $\mathbf{H}(3,5,L)$ with components (see (\ref{eq:sc_mat}))

\vspace{-0.25cm}


\begin{align}	
	\label{eq:H352}
	\mathbf{H}_0&=
		\left[
			\begin{array}{ccccc}
				\mathbf{0} & \mathbf{I}  & \mathbf{I}  & \mathbf{I}  & \mathbf{0}  \\
				\mathbf{0} & \mathbf{0} & \mathbf{0} & \mathbf{\sigma^3} & \mathbf{\sigma^4} \\
				\mathbf{I} & \mathbf{\sigma^2} & \mathbf{0} & \mathbf{0} & \mathbf{\sigma^3}  
			\end{array}
		\right], \\	
		\mathbf{H}_1&=
		\left[
			\begin{array}{ccccc}
				\mathbf{I} & \mathbf{0} & \mathbf{0}  & \mathbf{0}  & \mathbf{I}  \\
				\mathbf{I} & \mathbf{\sigma} & \mathbf{\sigma^2} & \mathbf{0} & \mathbf{0} \\
				\mathbf{0} & \mathbf{0} & \mathbf{\sigma^4} & \mathbf{\sigma} & \mathbf{0}  
			\end{array}
		\right],
\end{align}

\noindent and corresponding global code $\mathscr{C}$. Similarly, $\mathbf{H}_1(4,5)$ from (\ref{eq:3_mats}) can be used to generate the nested AB-SC-LDPC matrix $\mathbf{H}_1(4,5,L)$, using the example edge-spreading matrix  

\vspace{-0.25cm}
\begin{align}
\label{eq:B11}
	\mathbf{B}_{1,1}=\left[
		\begin{array}{ccccc}
			1 & 0 & 0 & 0 & 1 \\
			1 & 1 & 1 & 0 & 0 \\
			0 & 0 & 1 & 1 & 0 \\
			\color{blu}{1} & \color{blu}{0} & \color{blu}{0} & \color{blu}{1} & \color{blu}{0} \\
		\end{array}
	\right]\in \mathbb{F}_2^{4\times p},
\end{align}

\vspace{-0.25cm}
\noindent with components 

\vspace{-0.25cm}

\begin{align}
\label{eq:H1452}	
	\mathbf{H}_0&=
		\left[
			\begin{array}{ccccc}
				\mathbf{0} & \mathbf{I}  & \mathbf{I}  & \mathbf{I}  & \mathbf{0}  \\
				\mathbf{0} & \mathbf{0} & \mathbf{0} & \mathbf{\sigma^3} & \mathbf{\sigma^4} \\
				\mathbf{I} & \mathbf{\sigma^2} & \mathbf{0} & \mathbf{0} & \mathbf{\sigma^3}  \\
				\color{blu}{\mathbf{0}} & \color{blu}{\mathbf{\sigma^3}} & \color{blu}{\mathbf{\sigma}} & \color{blu}{\mathbf{0}} & \color{blu}{\mathbf{\sigma^2}}  
			\end{array}
		\right], \\
		\mathbf{H}_1&=
		\left[
			\begin{array}{ccccc}
				\mathbf{I} & \mathbf{0} & \mathbf{0}  & \mathbf{0}  & \mathbf{I}  \\
				\mathbf{I} & \mathbf{\sigma} & \mathbf{\sigma^2} & \mathbf{0} & \mathbf{0} \\
				\mathbf{0} & \mathbf{0} & \mathbf{\sigma^4} & \mathbf{\sigma} & \mathbf{0} \\
				\color{blu}{\mathbf{I}} & \color{blu}{\mathbf{0}} & \color{blu}{\mathbf{0}} & \color{blu}{\mathbf{\sigma^4}} & \color{blu}{\mathbf{0}} 
			\end{array}
		\right],
\end{align}

\noindent and corresponding nested code $\mathscr{C}_1$. Note that $\mathbf{H}_{1}(4,5,2)$ contains the (global) $\mathbf{H}(3,5,2)$ matrix, shown in black. The sub-matrix $\mathbf{H}_{2}(4,5,2)$ is constructed in a similar fashion from the spreading matrix $\mathbf{B}_{2,1}\in \mathbb{F}_2^{4\times p}$. \hfill\(\Box\)

\end{exmp}

\noindent Note that the order of optimization determines how good the sub-codes or the global code will be in terms of the number of harmful ABSs. We will see later in Section \ref{sec:Opt_Alc} that for column weight-$3$, $4$, or $5$ nested AB-SC-LDPC codes, where harmful ABSs are known to contain 6-cycles, ALC can be employed effectively to speed-up the optimization procedure. 

The optimization of nested AB-SC-LDPC codes may be performed in several ways. In any method, the spreading of all row groups $\mathcal{S}=\{0,1,\ldots,\gamma-1\}$ must be determined, but the results will vary depending on the order of the optimization. In the following, we consider two general exemplary methods. An example and {a} discussion will follow in Section \ref{sec:constrained_cons}. \\

\vspace{-0.5cm}
\noindent \textbf{Method 1}: First, optimize the $\mathbf{B}_m$ matrix, containing row groups $\mathcal{S}_0=\{0,1,2\}$, to construct the global matrix $\mathbf{H}(3,p,L)$ from $\mathbf{H}(3,p)$. Then, pick an ordering of $\{1,2,\ldots,M\}$, denoted $\mathbf{t}=(t_1,t_2,\ldots,t_M)$, and by incorporating the constraints given by $\mathbf{B}_m$, sequentially optimize for  $i= 1,2,\ldots,M$, the remaining rows of each $\mathbf{B}_{t_i,m}$ matrix containing row groups $\mathcal{S}_{t_i}\subseteq \mathcal{S}$, \emph{i.e.}, those row groups $\mathcal{S}_{t_i}\cap (\mathcal{S}\setminus \cup_{j=1}^{i-1}\mathcal{S}_{t_j})$ in $\mathcal{S}_{t_i}$ that have not yet been determined, to construct $\mathbf{H}_{t_i}(\omega_i,p,L)$ from $\mathbf{H}_{t_i}(\omega_i,p)$. \hspace{-0.15cm}\footnote{Note that the order $\mathbf{t}$ of sub-codes can be changed depending on priority. If codes have no overlapping row groups (other than $S_0$), the order can be chosen arbitrarily with no effect in results.} 

\vspace{-0cm}
\noindent \textbf{Method 2}: Again, pick an ordering $\mathbf{t}=(t_1,t_2,\ldots,t_M)$ of $\{1,2,\ldots,M\}$. Here, we first optimize the nested $\mathbf{B}_{t_1,m}$ matrix, with row groups $\mathcal{S}_{t_1}=\{0,1,2,\ldots,w_{t_1}-2,w_{t_1}-1+h_{t_1}\}$ to construct $\mathbf{H}_{t_1}(\omega_{t_1},p,L)$ from $\mathbf{H}_{t_1}(\omega_{t_1},p)$ (note that this nested matrix contains the global matrix corresponding to $\mathcal{S}_0=\{0,1,2\}$, and thus the global code is not independently optimized). Given the constraints of the newly generated $\mathbf{B}_{t_1,m}$ matrix, sequentially optimize the matrix $\mathbf{B}_{t_i,m}$, $i=2,3,\ldots,M$, by determining the spreading of any row groups in $\mathcal{S}_{t_i}\cap (\mathcal{S}\setminus \cup_{j=1}^{i-1}\mathcal{S}_{t_j})$, to construct $\mathbf{H}_{t_i}(\omega_{t_i},p,L)$ from $\mathbf{H}_{t_i}(\omega_{t_i},p)$.
\vspace{-0.3cm}


\subsection{Constrained Optimization}
\label{sec:constrained_cons}

Suppose that $\mathbf{B}_{t_1,m}$ is optimized first (\emph{i.e.}, without constraints) using Method 2. This unconstrained optimization determines any edge-spreading matrix $\mathbf{B}_{t_i,m}$, for $i\neq 1$, with a set of row indices $\mathcal{S}_{t_i}\subset \mathcal{S}_{t_1}$, and partially determines any edge-spreading matrix $\mathbf{B}_{t_i,m}$, $i\in \{2,3,\ldots,M\}$ with a set of row indices $\mathcal{S}_{t_i}\not\subset \mathcal{S}_{t_1}$, but $\mathcal{S}_{t_i}\cap \mathcal{S}_{t_1}\neq \emptyset$. The row(s) of $\mathbf{B}_{t_i,m}$ that are not optimized yet are determined via constrained optimization.  Although the order of constrained optimization may be chosen in an ad hoc manner, the freedom in choosing the entries of $\mathbf{B}_{t_i,m}$ may diminish as more edge-spreading matrices are determined prior to it. An example is provided in the following to demonstrate the constrained optimization process (which is applicable to both Methods 1 and 2, but shown only for Method 2).

\begin{exmp}
Fig. \ref{fig:constraint} shows the row group indices of four nested AB-LDPC-BC sub-codes using four colored boxes. The indices enclosed within the blue, red, green and black boxes, correspond to respective row groups $\mathcal{S}_1=\{0,1,\ldots,\omega-1\}$, $\mathcal{S}_2=\{0,1,\ldots,\omega,\gamma-3\}$, $\mathcal{S}_3=\{0,1,\ldots,\gamma-4,\gamma-2\}$, and $\mathcal{S}_4=\{0,1,\ldots,\gamma-3,\gamma-1\}$ of $\mathbf{H}(\gamma,p)$, and they form the respective sub-matrices $\mathbf{H}_1(\omega,p)$, $\mathbf{H}_2(\omega+2,p)$, $\mathbf{H}_3(\gamma-2,p)$, and $\mathbf{H}_4(\gamma-1,p)$ of $\mathbf{H}(\gamma,p)$, with respective column weights $\omega$, $\omega+2$, $\gamma-2$, and $\gamma-1$, and corresponding respective nested sub-codes $\mathscr{C}_1$, $\mathscr{C}_2$, $\mathscr{C}_3$, and $\mathscr{C}_4$, where $\mathscr{C}_2,\mathscr{C}_3 \subset \mathscr{C}_1$, $\mathscr{C}_4 \subset \mathscr{C}_2$, and $3< \omega< \gamma$. To construct an associated AB-SC-LDPC code, we select an optimization order, suppose $\mathbf{t}=(2,1,3,4)$. First, spreading matrix $\mathbf{B}_{2,m}$ is determined. Note that this determines $\mathbf{B}_{1,m}$ since $\mathcal{S}_1\subset \mathcal{S}_2$. Next, $\mathbf{B}_{3,m}$ is partially determined, but row groups $\omega+1,\omega+2,\ldots,\gamma-4,\gamma-2$ must now be optimized. Finally, $\mathbf{B}_{4,m}$ is partially determined by both $\mathbf{B}_{2,m}$ and $\mathbf{B}_{3,m}$, and the only remaining row group to optimize is $\gamma-1$. Note that, here, the global code $\mathscr{C}$ corresponds to the sub-matrix $\mathbf{H}(3,p)$ of $\mathbf{H}(\gamma,p)$ with row groups $0,1$, and $2$. The edge-spreading matrices are now used to construct $\mathbf{H}_1(\omega,p,L)$, $\mathbf{H}_2(\omega+2,p,L)$, $\mathbf{H}_3(\gamma-2,p,L)$ and $\mathbf{H}_4(\gamma-1,p,L)$, respectively, as described in Section \ref{sec:guide}. 

\begin{figure}[h]
\centering
\includegraphics[scale=0.65]{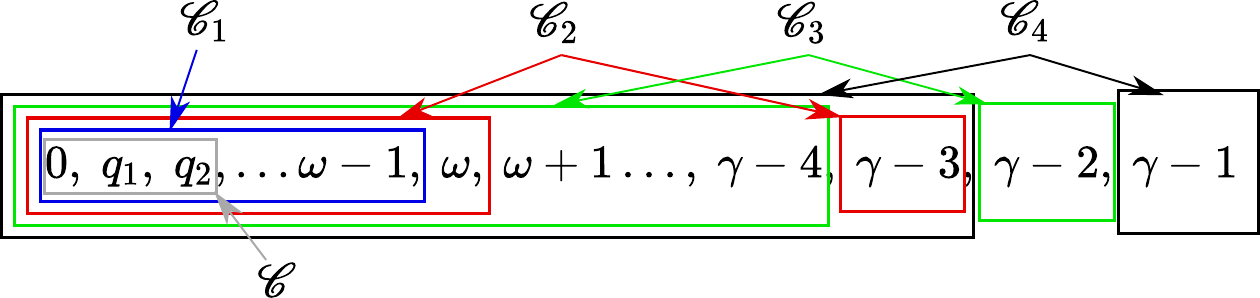}
\caption{An example of row group indices for four nested AB-LDPC-BC parity-check matrices shown using blue, red, green, and black boxes corresponding to nested sub-codes $\mathscr{C}_1$, $\mathscr{C}_2$, $\mathscr{C}_3$, and $\mathscr{C}_4$, respectively. The grey box contains the row group indices of the global code  $\mathscr{C}$.}
\vspace{-0.5cm}
\label{fig:constraint}
\end{figure}

Recall that the goal of our nested code optimization is to minimize the number of dominant ABSs in the sub-code's Tanner graph. In terms of sub-codes, the constrained optimization order in this example would give the most freedom to code $\mathscr{C}_2$, and we would expect that code to have relatively fewer harmful ABSs. On the other hand, {the} nested code $\mathscr{C}_4$ does not have much freedom since it contains only one undetermined row group prior to optimization, and thus it may exhibit relatively more harmful ABSs. Depending on the application, $\mathbf{t}$ should be chosen carefully. \hfill\(\Box\)
\end{exmp}

\subsection{Terminal Lift}
QC codes are well known to facilitate hardware implementation \cite{CDFKMS14}. To simplify our code search for good nested code designs, we choose to apply a circulant-based graph lifting after the steps described in Sections \ref{sec:general_cons} and \ref{sec:constrained_cons}, which we refer to as a \emph{terminal lift} with lifting factor $J$. This results in quasi-cyclic LDPC-BCs (QC-LDPC-BCs) and QC-SC-LDPC codes. {A similar lifting procedure has been shown to be very effective for constructing QC codes \cite{WDY13}.}  The parity-check matrices obtained by applying a terminal lift to $\mathbf{H}_i(\omega_i,p,L)$ and $\mathbf{H}(3,p,L)$ are denoted as $\mathbf{H}_i(\omega_i,p,L,J)\in \mathbb{F}_2^{\omega_i pJ(L+m) \times JLp^2}$, $i=1,2,\ldots,M$, and $\mathbf{H}(3,p,L,J)\in \mathbb{F}_2^{3pJ(L+m)\, \times\, JLp^2}$, respectively, with constraint length $\nu'=J\nu$. A terminal lift serves two purposes: first it helps us generate sufficiently long nested codes to achieve good performance for typical applications, and second, we are able to further reduce the multiplicity of, or even eliminate, any residual ABSs in the nested Tanner graphs of $\mathbf{H}_i(\omega_i,p,L)$ and $\mathbf{H}(3,p,L)$ that remain after the edge-spreading construction. Note that, for comparison, we will also construct a terminally lifted LDPC-BC matrix, denoted as $\mathbf{H}(\gamma,p,J(m+1))\in \mathbb{F}_2^{\gamma pJ(m+1) \times J(m+1)p^2}$, by lifting the block matrix $\mathbf{H}(\gamma,p)$ by a larger lifting factor $J(m+1)$. By incorporating $m+1$ in the lifting factor we ensure that the length (number of columns) of the BC matrix is identical to the constraint length of the SC matrix. This will be helpful for comparing the performances of {QC}-LDPC-BCs and {QC}-SC-LDPC codes in Section \ref{sec:results}. 


\subsection{6-cycles and Dominant ABSs in AB-LDPC Codes}
\label{sec:ABSs}

\noindent We now state some useful results concerning dominant objects of AB-SC-LDPC codes. 

\begin{rem}
\label{rem:elim_64}
 From Remarks \ref{rem:6cyc_33ABS} and \ref{rem:64abs} it is easy to note that by eliminating all 6-cycles via spatial coupling, we can also eliminate all the dominant ABSs in $\gamma=3,4,5$ AB-SC-LDPC codes.
\end{rem}

\noindent Note that a 6-cycle in an AB-SC-LDPC matrix spans $3$ row groups and it spans at most $m+1$ contiguous column blocks of (\ref{eq:sc_mat}) \cite{MR17}. Let $\mu_e$, $e=1,2,\ldots,m+1$, represent the total number of 6-cycles present in precisely $e$ contiguous column blocks of (\ref{eq:sc_mat}), and let $\mu$ be the total number of 6-cycles in $\mathbf{H}(\gamma, p, L)$. From the repeated structure of $\mathbf{H}(\gamma, p, L)$ it follows that the total number of 6-cycles in $\mathbf{H}(\gamma, p, L)$ is $\mu=\sum_{e=1}^{m+1} (L-e+1)\mu_e$ \cite{MR17}.

\subsection{Adapted Line Counting (ALC)}
\label{sec:ALC}

\vspace{-0cm}
The line-counting method discussed in Section \ref{sec:LC} is restricted to the enumeration of 6-cycles in $\mathbf{H}(3,p,L)$, \emph{i.e.}, only to row groups $0,1$, and $2$. To overcome this shortcoming, we propose an ALC scheme to enable 6-cycle enumeration  in general row group triples $\{0,q_2,q_3\}$ of the larger parity-check matrix $\mathbf{\hat{H}}$. 


\subsubsection{Choosing Column weight-$3$ Sub-matrices for ALC}
\label{sec:sub_sel}

To optimize the nested construction, we need to optimize each row group. Our method involves optimization with row groups of weight-$3$. We begin by enumerating the required sub-matrices involved in the optimization. Let $\zeta=\{\mathbf{H}_i^{(1)}(3,p),\mathbf{H}_i^{(2)}(3,p),\ldots,\mathbf{H}_i^{(Z)}(3,p)\}$ be the set of all $Z=$ $\omega_i \choose 3$$-$$\omega_i-1 \choose 3$ column weight-$3$ sub-matrices of $\mathbf{H}_i(\omega_i,p)$ having the row group $0$ in common. We first select a subset  $\zeta'=\{\mathbf{H}_i^{(z_1)}(3,p),\mathbf{H}_i^{(z_2)}(3,p),\ldots,\mathbf{H}_i^{(z_{Z'})}(3,p)\}\subseteq \zeta$, $z_1,z_2,\ldots$,$z_{Z'}\in \{1,2,\ldots,Z\}$, $Z'=\floor{(\omega_i-1)/2}$, where each matrix in $\zeta'$ consists of a common row group $0$ and a distinct row group pair $(q,q')$ from all other elements of $\zeta'$. If $\omega_i-1$ is odd, there is one more row group that has not appeared in any matrix in $\zeta'$, and we need to select an additional matrix $\mathbf{H}_i^{(z_{Z'+1})}(3,p)$ in $\zeta\setminus \zeta'$ with row groups $\{0,q^{\dprime},q'''\}$, where either the row group $q^{\dprime}$ or $q'''$ is not shared with a matrix in $\zeta'$. The number of possible $\mathbf{H}_i^{(z_{Z'+1})}(3,p)$ matrices is $Z^{\dprime}=(\omega_i-1)\text{ mod }2$, and it follows that we must select $T=Z'+Z^{\dprime}$ distinct column weight-$3$ sub-matrices in $\mathbf{H}_i(\omega_i,p)$ that have a common row group $0$ in order to include all of the row groups of $\mathbf{H}_i(\omega_i,p)$. Therefore, to optimize the nested SC matrix $\mathbf{H}_i(\omega_i,p,L)$ corresponding to $\mathbf{H}_i(\omega_i,p)$, it is sufficient to enumerate 6-cycles in the $T$ sub-matrices of $\zeta$ detailed above using ALC (details to follow in Section \ref{sec:Opt_Alc}).

\begin{exmp}
{We provide an example for choosing the required column weight-$3$ sub-matrices in a nested matrix for ALC, with nested column weights $\omega_i=4$ and $5$. 
In the first case, the $\mathbf{H}_i(4,p)$ matrix has row groups $0,1,2$, and $3$. The set $\zeta$ contains $Z=2$ column weight-$3$ sub-matrices with a common row group $0$: $\zeta=\{\mathbf{H}_i^{(1)}(3,p),\mathbf{H}_i^{(2)}(3,p)\}$, with respective row groups $\{0,1,2\}$ and $\{0,1,3\}$ of $\mathbf{H}_i(4,p)$. Hence, we select $T=Z'+Z^{\dprime}=1+1=2$ matrices to include all row groups, \emph{e.g.}, $\zeta'=\{\mathbf{H}_i^{(1)}(3,p)\}$, and we select the additional matrix $\mathbf{H}_i^{(2)}(3,p)$ from $\zeta\setminus \zeta'$.
In the second case, the $\mathbf{H}_i(5,p)$ matrix has row groups $0,1,2,3$, and $4$. The set $\zeta$ contains $Z=6$ column weight-$3$ sub-matrices with a common row group $0$: $\zeta=\{\mathbf{H}_i^{(1)}(3,p),\mathbf{H}_i^{(2)}(3,p),\mathbf{H}_i^{(3)}(3,p),$ $\mathbf{H}_i^{(4)}(3,p),\mathbf{H}_i^{(5)}(3,p),\mathbf{H}_i^{(6)}(3,p)\}$, with respective row groups $\{0,1,2\}$, $\{0,1,3\}$, $\{0,1,4\}$, $\{0,2,3\}$, $\{0,2,4\}$, and $\{0,3,4\}$ of $\mathbf{H}_i(5,p)$. We must select $T=Z'+Z^{\dprime}=2+0=2$ matrices to include all row groups, \emph{e.g.}, $\zeta'=\{\mathbf{H}_i^{(1)}(3,p),\mathbf{H}_i^{(6)}(3,p)\}$. 
} \hfill\(\Box\)

\end{exmp}

\subsubsection{Description of ALC}
\label{sec:ALC_math}
Recall from Section \ref{sec:LC} that $\mathcal{R}$ is contained in a column weight-$3$ sub-matrix of an AB parity-check matrix, and $c_1,c_2,c_3$ are the column indices of a 6-cycle in $\mathcal{R}$. We begin by extending (\ref{eq:ineq1}) to a general column weight-$3$ matrix containing row group $0$. 

\begin{lem}
\label{lem:ALC}
	For ALC, the range of $c_3$ in $\mathcal{R}$ is expressed via $c_1$ and $c_2$ as
	\begin{align}
	\label{eq:ineq2}
	\begin{split}	
		\left(1-\frac{q_3}{q_2}\right)\alpha p+ \frac{\lambda p^2}{q_2}\leq c_2-\frac{q_3}{q_2}c_1<\left(1-\frac{q_3}{q_2}\right)\beta p+ \frac{\lambda p^2}{q_2}, 		
	\end{split}
	\end{align}
	
	\vspace{-0.1cm}
	\noindent where $c_2>c_1$, $\lambda\in\{2-2p,\ldots,2p-2\}$, $q_2,q_3\in\{1,\ldots,\gamma-1\}$ and $q_2\neq q_3$.
\end{lem}

\noindent The proof is given in Appendix \ref{app:range}. Note that (\ref{eq:ineq1}) is a special case of (\ref{eq:ineq2}) because, for parameters $(\lambda,q_2,q_3)$ taking on the values $(0,2,1)$, $(1,2,1)$, $(1,1,2)$, and $(0,1,2)$, we recover all inequalities in (\ref{eq:ineq1}). 

\begin{prop}
\label{prop:alc}
\emph{Based on the principles of Cartesian geometry, the total number of 6-cycles in $\mathcal{R}$, $\mathcal{N}_{\mathcal{R}}$, is given by}

\vspace{-0.5cm}
\begin{equation}
\label{eq:ALC}
\mathcal{N}_{\mathcal{R}}=
\begin{cases}
&\sqrt{\frac{(\sigma_{2x}-\sigma_{1x})^{2}+(\sigma_{2y}-\sigma_{1y})^{2}}{2}}, \\ &\text{ if some boundary conditions\footnotemark \hspace{0.01cm} hold}, \\
&0, \text{otherwise},
\end{cases}
\end{equation}

\footnotetext{These boundary conditions are stated in (\ref{eq:conditions}) but omitted here for brevity.}


\noindent \emph{where the boundary conditions in (\ref{eq:ALC}) (stated in (\ref{eq:conditions})) are related to the $(c_1,c_2)$ plane and the line $c_2-c_1=np$, $n=1,2,\ldots,w_4-w_1-1$, whose length is determined by the points $(\sigma_{1x},\sigma_{1y})$, $(\sigma_{2x},\sigma_{2y})$ within these boundaries.}
\vspace{-0.15cm}
\end{prop}

\noindent The proof of this proposition, including the boundary conditions, is given in Appendix \ref{app:ALC}. {In general, suppose a $k$-cycle has a row $r_i$, $i\in \{1,2,\ldots,\frac{k}{2}-1\}$, in the $q_0$ row group. If this row is connected to columns $c_i$ and $c_{i'}$, $i,i'\in \{1,2,\ldots,\frac{k}{2}-1\}$ and $i\neq i'$, we can obtain a line $c_i-c_{i'}=np$ on the $(c_1,c_2,\ldots,c_{\frac{k}{2}-1})$ plane, which passes through a $(\frac{k}{2}-1)$-dimensional region $\mathcal{L}$, enclosed by the boundaries generated by the $k$-cycle column values. Thus, any integer point lying on this line will represent that $k$-cycle, allowing line-counting to be further adapted to detect and enumerate cycles of length longer than six.}

Let $N$ denote the number of regions in an AB-LDPC matrix required by ALC to compute all the 6-cycles, which is independent of $p$ but may depend on $m$. In case of AB-LDPC-BCs, $N=1$ as $\mathcal{R}$ is the entire parity-check matrix. However, $N>1$ for column weight-$3$ AB-SC-LDPC parity-check matrices since there are multiple matrix regions with distinct row group triples and no two regions share a 6-cycle. {For code optimization, these triples need to be considered only for coupling length $L=m+1$, since a 6-cycle does not span more than $m+1$ contiguous column blocks of a column weight-$3$ AB-SC-LDPC parity-check matrix (see Section \ref{sec:ABSs}).}

\begin{prop} 
\label{prop:Nubound}
{$N$ is upper bounded by $(3m+2)\bigg(\binom{2(m+1)}{2}-2\binom{m+1}{2}\bigg)$.}
\end{prop}

\noindent {The proof is given in Appendix \ref{app:Nubound}.} An example of determining $N$ in the case of a column weight-$3$ and memory $m=1$ AB-SC-LDPC matrix is provided in the following.    

\begin{exmp} \emph{Determining $N$ for memory $m=1$}: Note that in this case, it is necessary to optimize only two contiguous column blocks of a column weight $\omega=3$ (time-invariant) AB-SC-LDPC matrix $\mathbf{H}_i^{(z)}(3,p,L)$. Hence, it is sufficient to consider coupling length $L=2$. Note that, for $L=2$, there are three copies of the distinct row groups $0,q_2,q_3$ from $\mathbf{H}(\gamma,p)$ in $\mathbf{H}_i^{(z)}(3,p,2)$. Recall that we locate a copy of a row group by the ordered pair $(y,q)$, {where, $y\in \{0,1,\ldots,L+m-1=2\}$ is the row block index} and $q$ is the row group index. Here, we slightly abuse notation and identify the row group index as $q\in\{0,q_2,q_3\}$, where it is understood that $q_2$ identifies the second row group in a row block of $\mathbf{H}_i^{(z)}(3,p,2)$ and $q_3$ identifies the third, respectively. We choose to do this because the values of $q_2$ and $q_3$ (corresponding to the row groups and circulants of $\mathbf{H}(\gamma,p)$) will be used in ALC as described in Section \ref{sec:ALC_math}.   

There are two possible region configurations. The first is a single column block $[\mathbf{H}_0^\top \text{ } \mathbf{H}_1^\top]^\top$ (\emph{i.e.,} matrix (\ref{eq:sc_mat}) with $L=1$). Here, a 6-cycle may exist in any of the $8$ regions $\mathcal{R}_1,\mathcal{R}_2,\ldots,\mathcal{R}_8$ with row group triples $\{(0,0),(0,q_2),(0,q_3)\},\{(0,0),(0,q_2),(1,q_3)\}$, $\{(0,0),(0,q_3),(1,q_2)\},\{(0,0),(1,q_2),(1,q_3)\},\{(1,0),$ $(0,q_2),(0,q_3)\},\{(1,0),(0,q_2),(1,q_3)\},\{(1,0),(0,q_3),(1,q_2)\}$, and $\{(1,0),(1,q_2),(1,q_3)\}$, respectively. Each of these regions span $p$ column groups and can be viewed as a block matrix similar to $\mathbf{H}(3,p)$ but with one or more all-zero circulant matrices. 


A 6-cycle may also span an ``L'' shaped matrix region configuration 

\vspace{-0.5cm} 
\[\begin{bmatrix}
\mathbf{H}_0 & \\
\mathbf{H}_1 & \mathbf{H}_0 \\
\end{bmatrix} \text { or } \begin{bmatrix}
\mathbf{H}_1 & \mathbf{H}_0\\
 & \mathbf{H}_1 \\
\end{bmatrix}.\] 

\noindent Seven regions of this category exist, denoted as $\mathcal{R}_9,\mathcal{R}_{10},\ldots,\mathcal{R}_{15}$, with row group triples \\ {$\{(0,0),(1,q_2),(1,q_3)\},\{(0,q_2),(1,0),(1,q_3)\},\{(0,q_3),(1,0),$ $(1,q_2)\},\{(1,0),(1,q_2),(1,q_3)\},\{(1,0),(1,q_2),(2,q_3)\}$, $\{(1,0),(1,q_3),(2,q_2)\}$, and $\{(1,q_2),(1,q_3),(2,0)\}$}, respectively, with $2p$ column groups in each region. The total number of 6-cycles is obtained by enumerating the number of 6-cycles in each of the $N=15$ regions as $\mathcal{R}_1,\mathcal{R}_2,\ldots,\mathcal{R}_{15}$. \hfill\(\Box\)
\label{exmp:6_cycle_enum}
\end{exmp}

Note that, in a column weight-$3$ AB-LDPC matrix region $\mathcal{R}$, there are at most $p^2-p$ block cycles \hspace{-0.15cm}\footnote{There are $p^2(p-1)$ 6-cycles in a column weight-$3$ AB-LPDC-BC matrix \cite{AREKD16}, where there are $p^2-p$ block cycles of length $6$ with each block containing $p$ 6-cycles.} of length $6$, and each of them generate a separate ``block-cycle region'' $\mathscr{L}_y$, $y\in \{1,2,\ldots,p^2-p\}$, on the $(c_1,c_2)$ plane which the line $c_2-c_1=np$ may intersect. The location of $\mathscr{L}_y$ is determined by the range of values of $c_1$, $c_2$, and $c_3$ associated with the corresponding block 6-cycle. If $\mathcal{R}$ does not contain all-zero matrices, \emph{i.e.,} $\mathcal{R}$ is the $\mathbf{H}(3,p)$ matrix, then regions $\mathscr{L}_1,\mathscr{L}_2,\ldots,\mathscr{L}_{p_2-p}$ combine to form a single consecutive region $\mathscr{L}$ on the $(c_1,c_2)$ plane. Therefore, the total number of 6-cycles in $\mathcal{R}$ will be identical to the length of the line $c_2-c_1=np$ passing through $\mathscr{L}$, and hence, only a single line computation will be necessary to compute $\mathcal{N}_{\mathcal{R}}$. However, if $\mathcal{R}$ contains one or more all-zero matrices, \emph{i.e.,} $\mathcal{R}$ is one of the matrix regions $\mathcal{R}_1,\mathcal{R}_2,\ldots,\mathcal{R}_{15}$ discussed in Example \ref{exmp:6_cycle_enum}, then, one or more region(s) from $\mathscr{L}_1,\mathscr{L}_2,\ldots,\mathscr{L}_{p_2-p}$ will be eliminated, depending on which block cycle(s) of  $\mathcal{R}$ is/are eliminated due to the presence of the all-zero matrices(s). Consequently, in this case, $\mathcal{N}_{\mathcal{R}}$ will be identical to the total number of integer points lying on the line $c_2-c_1=np$ within the existing block-cycle regions which may not be consecutive anymore. As a result, multiple line computations may be necessary to determine $\mathcal{N}_{\mathcal{R}}$, depending on how many (disjoint) block-cycle regions the line $c_2-c_1=np$ intersects.

\subsubsection{Complexity} 
The conventional cycle counting method discussed in \cite{LiLin15} has computational complexity of $\mathcal{O}(gE^2/p)=\mathcal{O}(g\gamma^2 p^3)$ for AB-LDPC-BCs, where $g$ is the girth and $E=\gamma p^2$ is the number of edges in the graph, respectively.

Since a column weight-$3$ matrix region $\mathcal{R}$ generates $p^2-p$ block-cycle regions on the $(c_1,c_2)$ plane, the number of line computations necessary for determining $\mathcal{N}_{\mathcal{R}}$ via ALC scales with $\mathcal{O}(p^2)$ since the number of arithmetic operations involved in a single line computation (corresponding to a single block-cycle region) is independent of the code parameters. Thus, due to a significantly reduced 6-cycle enumeration complexity compared to standard cycle counting algorithms, ALC is more desirable for the optimization of nested codes.

\section{AB-SC-LDPC Nested Code Optimization Procedure}
\label{sec:Opt}

In this section, we discuss the optimization strategy for {nested AB-SC-LDPC codes} in example scenarios where the nested codes have column weights $\omega_i=4$ or $5$, and the global code has column weight-$3$. First, a numerical optimization scheme is employed to determine the entries of the edge-spreading matrices $\mathbf{B}_{i,m}$ and $\mathbf{B}_m$ to construct parity-check matrices $\mathbf{H}_i(\omega_i,p,L)$ and $\mathbf{H}(\omega=3,p,L)$ from $\mathbf{H}_i(\omega_i,p)$ and $\mathbf{H}(\omega=3,p)$, respectively. ALC is used at each optimization step to compute (and minimize) the number of 6-cycles in the nested AB-SC-LDPC parity-check matrices. Finally, a terminal lift is applied to further reduce residual ABSs in the Tanner graphs of optimized $\mathbf{H}_i(\omega_i,p,L)$ and $\mathbf{H}(3,p,L)$ matrices.

\subsection{ALC-Based Guided Search}
\label{sec:guide}

An overview of the search mechanism for determining the entries of the spreading matrices is provided in the following. For ease of notation, we refer to a given AB-LDPC-BC parity-check matrix $\mathbf{H}(3,p)$ or $\mathbf{H}_i(\omega_i,p)$ as $\mathbf{H}$, the corresponding undetermined edge-spreading matrix $\mathbf{B}_m$ or $\mathbf{B}_{i,m}$ as $\mathbf{B}$, and the resulting matrix $\mathbf{H}(3,p,L)$ or $\mathbf{H}_i(\omega_i,p,L)$ obtained by edge-spreading $\mathbf{H}$ according to $\mathbf{B}$ as $\mathbf{H}_{sc}$. Entries in $\mathbf{B}$ that have been determined by optimization of another nested code (see the constrained optimization discussion in Section \ref{sec:constrained_cons}) are copied into $\mathbf{B}$ and cannot be changed, the remaining entries are initially set to zero (\emph{i.e.,} the corresponding circulants exist in sub-matrix $\mathbf{H}_0$ of (\ref{eq:sc_mat})) and designated to be ``unfixed''. In each search iteration $\ell$, either an unfixed entry of $\mathbf{B}$ is determined, or ``fixed'', referred to as forward-search, or a fixed entry of $\mathbf{B}$ is erased and re-evaluated, referred to as back-tracking. In a forward-search step, an integer from the set $\mathcal{S}'=\{0,1,\ldots,m-1\}$ is assigned to an unfixed entry $(i,j)$ of $\mathbf{B}$, and the number of 6-cycles in the corresponding $\mathbf{H}_{sc}$ matrix is computed via ALC. This procedure is repeated for all the remaining elements in $\mathcal{S}'$, and the one that contributes to the least number of 6-cycles in $\mathbf{H}_{sc}$ is chosen for entry $(i,j)$ of $\mathbf{B}$ and designated fixed. Note that there may be more than one value in $\mathcal{S}'$ which corresponds to the same number of 6-cycles for entry $(i,j)$. In this case, entry $(i,j)$ is set to be one of those values that minimizes the cycles from $\mathcal{S}'$ randomly, and the remaining candidates are stored for back-tracking. After each round $\ell$ of forward-search, the number of 6-cycles in $\mathbf{H}_{sc}$ is given by the integer $\rho_{\ell}$. If $\rho_{\ell}\geq\rho_{\ell-1}$, then a back-tracking step is applied. In this phase, the most recently assigned entry $(i,j)$ is designated unfixed, then the forward-search is applied and a new value is chosen for the unfixed entry randomly from the previous set of saved candidates for that entry. The optimization procedure terminates if no more back-tracking is possible, or after a maximum number of search iterations ($\ell_{\max}$) elapse, or if $\rho_{\ell}=0$, whichever occurs first. 





\vspace{-0.25cm}
\subsection{Construction of Optimized Nested AB-SC-LDPC Codes}  
\label{sec:Opt_Alc}

We outline the approach of Method 1 using ALC, which first optimizes $\mathbf{H}(3,p,L)$ and then the $\mathbf{H}_{t_i}(\omega_i,p,L)$ matrices based on the constraints given by $\mathbf{H}(3,p,L)$ and the ordering $\mathbf{t}$, as described in Section \ref{sec:general_cons}. The optimization requires the following three steps:

\begin{itemize}
	\item \textbf{Step 1:} Inputs to the ALC based optimization algorithm are the $\mathbf{H}(3,p)$ matrix and the undetermined $\mathbf{B}_m$ matrix. The guided search is allowed to run as described in Section \ref{sec:guide} until either the number of 6-cycles in $\mathbf{H}(3,p,L)$ obtained in an optimization iteration is $0$, or until the number of optimization iterations exceed a predetermined threshold $\ell_{\max}$, whichever occurs first. At the end of this step we obtain $\mathbf{B}_m$, and the spreading for row groups $0,1$, and $2$ is permanently fixed.
	\item \textbf{Step 2:} For a given sub-code index $t_i$, inputs to the ALC based optimization algorithm are $\mathbf{H}_{t_i}^{(z)}(3,p)$, consisting of row groups $\{0,q,q'\}$, and a matrix $\mathbf{B}_{t_i,m}^{(z)}\in \mathbb{F}_2^{3\times p}$ which contains the rows corresponding to row groups $\{0,q,q'\}$ of $\mathbf{B}_{t_i,m}$. Note that the row $0$ of $\mathbf{B}_{t_i,m}^{(z)}$ has been fixed already and cannot be changed; if the row(s) corresponding to the row groups $q$ and/or $q'$ of $\mathbf{B}_{t_i,m}^{(z)}$ are/is unfixed, then they/it will be fixed in this step. The guided search algorithm of Section \ref{sec:guide} is now allowed to run until either all the 6-cycles in $\mathbf{H}_{t_i}^{(z)}(3,p,L)$, obtained by edge-spreading $\mathbf{H}_{t_i}^{(z)}(3,p)$ using $\mathbf{B}_{t_i,m}^{(z)}$, are eliminated or the number of optimization iterations exceed a threshold $\ell_{\max}$, whichever occurs first. This optimization step is repeated for all possible values of $z$ from $\{2,3\ldots,T\}$. At the end of this step, we obtain the complete  $\mathbf{B}_{t_i,m}$ matrix. This step is repeated until all sub-codes $i=1,2,\ldots,M$ are set.
	\item \textbf{Step 3:} Finally, the $\mathbf{H}_i(\omega_i,p,L)$ and $\mathbf{H}(3,p,L)$ matrices are constructed by edge-spreading according to $\mathbf{B}_{i,m}$. 
\end{itemize}

\noindent Method 2 largely follows the approach in Method 1, but here $\mathbf{H}_i(\omega_i,p,L)$ is optimized first instead of $\mathbf{H}(3,p,L)$ (as described in Section \ref{sec:general_cons}). 


\begin{exmp}
\label{exmp:sc}

\vspace{-0.1cm}
\noindent We demonstrate ALC based optimization for Method 1 using the AB-LDPC-BC from Example \ref{exmp:block} with nested matrix $\mathbf{\hat{H}}(5,5)$, coupling length $L=2$, and memory $m=1$. In Step 1, we input an all-zero $3\times 5$ $\mathbf{B}_1$ matrix and return the optimized $\mathbf{B}_1$ matrix shown earlier in (\ref{eq:B_1}). This is used to generate the optimized global SC matrix $\mathbf{H}(3,5,2)$, whose components are shown in (\ref{eq:H352}), from $\mathbf{H}(3,5)$. In Step 2, we select $\mathbf{t}=(1,2)$ and first optimize the matrix $\mathbf{B}_{1,1}$ that will be used to spread the edges of $\mathbf{H}_1(4,5)$ from (\ref{eq:3_mats}) to construct the optimized (with constraints) nested AB-SC-LDPC code $\mathscr{C}_1$. \hspace{-0.3cm} \footnote{Note that, since row groups $0$, $1$, and $2$ are fixed in Step 1, $\mathbf{t}=(1,2)$, and $\mathbf{t}=(2,1)$ would have the same outcome.} Note that, for this code, $\mathbf{H}_1(4,5)$ consists of row groups $0,1,2$, and $3$, of which $0,1$, and $2$ (black rows) are fixed in Step 1. Here, $T=2$ and the two matrices $\mathbf{H}_1^{(z)}(3,p)$, $z=1,2$, can be chosen, say, to have row groups $\{0,1,2\}$ {and $\{0,2,3\}$,} respectively. Since $\mathbf{H}_1^{(1)}(3,p)$ is completely determined and there is only on row group {(with index $3$)} to fix, we require only one optimization ($z=2$). The inputs in Step 2 are the sub-matrix 

\vspace{-0.25cm}		
\begin{align}
\label{eq:H1235}
	     \mathbf{H}_1^{(2)}(3,5)=
		\left[
				\begin{array}{ccccc}
					\mathbf{I} & \mathbf{I}  & \mathbf{I}  & \mathbf{I}  & \mathbf{I} \\
					\mathbf{I} & \mathbf{\sigma^2} & \mathbf{\sigma^4} & \mathbf{\sigma} & \mathbf{\sigma^3} \\
					\color{blu}{\mathbf{I}} & \color{blu}{\mathbf{\sigma^3}} & \color{blu}{\mathbf{\sigma}} & \color{blu}{\mathbf{\sigma^4}} & \color{blu}{\mathbf{\sigma^2}} \\	
				\end{array}
			\right]
\end{align}

\vspace{-0cm}
\noindent of $\mathbf{H}_1(4,5)$ and the matrix $\mathbf{B}_{1,1}^{(2)}$, where the first two rows of $\mathbf{B}_{1,1}^{(2)}$, corresponding to row groups $0$ and {$2$}, were already determined in Step 1 as {rows $0$ and $2$} of $\mathbf{B}_1$ in (\ref{eq:B_1}), and the third row is set to all-zeros. The output is the optimized  matrix

\vspace{-0.25cm}		
\begin{align}
\label{eq:B11'}
	\mathbf{B}_{1,1}^{(2)}=\left[
		\begin{array}{ccccc}
			1 & 0 & 0 & 0 & 1 \\
			0 & 0 & 1 & 1 & 0 \\
			\color{blu}{\emph{1}} & \color{blu}{\emph{0}} & \color{blu}{\emph{0}} & \color{blu}{\emph{1}} & \color{blu}{\emph{0}} \\
		\end{array}
	\right].
\end{align}

\vspace{-0.25cm}
\noindent The blue row in italics of (\ref{eq:B11'}) represents the only row of $\mathbf{B}_{1,1}^{(2)}$ that has been optimized in this step. Now, combining $\mathbf{B}_1$ and $\mathbf{B}_{1,1}^{(2)}$, all the row groups $0,1,2$, and $3$ have been fixed and we obtain the optimized edge-spreading matrix $\mathbf{B}_{1,1}$ shown in (\ref{eq:B11}), that is used to construct the optimized nested SC matrix $\mathbf{H}_1(4,5,2)$, whose components are shown in (\ref{eq:H1452}). The sub-matrix $\mathbf{H}_{2}(4,5,2)$ is constructed in a similar fashion to complete Step 2. Finally, the nested matrices are constructed as described in Step 3. \hfill\(\Box\)


\end{exmp}

\subsection{Optimization Complexity}
Let $x\in \{1,2,\ldots,\omega'\}$, $\omega' \in \{3,\omega_i\}$, be the number of undetermined rows of a $\mathbf{B}$ matrix input to the ALC-based optimization scheme. For a given column weight-$3$ AB-LDPC-BC parity-check matrix $\mathbf{H}$, the scheme needs to search over at most $(m+1)^{xp}$ possible choices of the $xp$ entries belonging to the undetermined rows of $\mathbf{B}$, to determine the choices that minimize/eliminate the number of 6-cycles in $\mathbf{H}_{sc}$. Moreover, during a search, ALC is invoked for enumerating 6-cycles in $\mathbf{H}_{sc}$. Recall that $N$ column weight-$3$ matrix regions are taken into account by ALC to compute all the 6-cycles in $\mathbf{H}_{sc}$, and therefore at most $p^2-p$ line computations may be necessary for each region. Therefore, the complexity of our nested code optimization scheme for a given $\mathbf{B}$ is $\mathcal{O}(Np^2(m+1)^{x p})$. {Due to the high optimization complexity for large $p$, it may be more beneficial to choose a small value of $p$ and then subsequently apply terminal lifting to design longer SC codes.}

\vspace{-0cm}
\subsection{Terminal Lift}
\label{sec:Opt_term}

The terminal lifted matrices $\mathbf{H}_i(\omega_i,p,L,J)$, $i=1,2,\ldots,M$, and $\mathbf{H}(3,p,L,J)$ may now be obtained by lifting the non-zero (resp., zero) entries of $\mathbf{H}_i(\omega_i,p,L)$ and $\mathbf{H}(3,p,L)$ via randomly generated circulant (resp., all-zero) matrices of size $J\times J$. The goal of the terminal lift is to break as many remaining 6-cycles as possible in the Tanner graphs of $\mathbf{H}_i(\omega_i,p,L)$ and $\mathbf{H}(3,p,L)$. In this paper, we use an exhaustive search to  select circulants for the first $m+1$ contiguous column blocks of (\ref{eq:sc_mat}), searching until either all 6-cycles are eliminated or a maximum time duration is elapsed. Many other approaches could be considered, \emph{e.g.}, \cite{MSC14,BCBM19}. These permutations are repeated periodically for the remaining $L-(m+1)$ column blocks to form the complete  matrix. This is first performed for any $i$, \emph{i.e.,} on $\mathbf{H}_{i}(\omega_{i},p,L,J)$, and then repeated for the additional row groups in the parity-check matrices of the remaining nested codes $\mathscr{C}_1,\mathscr{C}_2,\ldots,\mathscr{C}_M$. Note that terminally lifted QC-LDPC-BC matrices, denoted $\mathbf{H}(3,p,J(m+1))$ and $\mathbf{H}_i(\omega_i,p,J(m+1))$, are obtained in similar fashion by lifting the non-zero (resp., zero) entries of $\mathbf{H}(3,p)$ and $\mathbf{H}_i(\omega_i,p)$ via circulant (resp., zero) matrices with dimension $J(m+1)\times J(m+1)$.

\section{Numerical Results}
\label{sec:results}

{We now demonstrate the effectiveness of the procedure outlined in Section \ref{sec:Opt} using nested AB-SC-LDPC matrices $\mathbf{H}(3,p,L)$, $\mathbf{H}_1(4,p,L)$, and $\mathbf{H}_4(5,p,L)$, with corresponding BC matrices $\mathbf{H}(3,p)$, $\mathbf{H}_1(4,p)$, and $\mathbf{H}_4(5,p)$ obtained from row groups $\{0,1,2\}$, $\{0,1,2,3\}$ and $\{0,1,2,3,4\}$, respectively, of a nested $\mathbf{\hat{H}}(6,p)$ AB-LDPC matrix. We provide 6-cycle enumeration results with syndrome former memories $m=1$ and $2$ in Section \ref{sec:results1}, and computer simulation results in Section \ref{sec:results2} with $m=2$.}

\vspace{-0.25cm}
\subsection{ALC Enumeration Results}
\label{sec:results1}

We first present results obtained only with the ALC-based optimization described in Section \ref{sec:Opt_Alc}, \emph{i.e.}, without a terminal lift. In the case of AB-SC-LDPC codes with memory $m$, we compute the asymptotic average number of 6-cycles per VN given by $A_m=\lim_{L \to \infty}\sum_{e=1}^{m+1} \frac{(L-e+1)\mu_e}{Lp^2}=\sum_{e=1}^{m+1}\frac{\mu_e}{p^2}$ \cite{MR17}. The asymptotic average results for nested AB-SC-LDPC matrices constructed via ALC based optimization with Methods 1 (M1) and Method 2 (M2), respectively, are shown in Table \ref{tab:tab1}. The maximum number of search iterations for the ALC based optimization scheme was set to $\ell_{\max}=10^4$. The asymptotic average results for random (Ran) {nested AB-SC-LDPC matrices} (non-optimized) are also presented in Table \ref{tab:tab1}, where randomly generated $\mathbf{B}_{m}$ and $\mathbf{B}_{i,m}$ matrices were selected. As another benchmark for comparison, we compute the asymptotic average number of 6-cycles per VN for AB-LDPC-BCs, given by $A^{(BC)}=\mu/p^2$, where $\mu$ is the total number of 6-cycles in the AB-LDPC-BC. The values of $A^{(BC)}$ in the matrix $\mathbf{H}(3,p)$ are computed as $12, 18$, and $30$, for $p=5,7,$ and $11$, respectively, for $\mathbf{H}_1(4,p)$ these values are $48, 72$, and $120$, respectively, {whereas for $\mathbf{H}_4(5,p)$, the $A^{(BC)}$ values are $60$ and $100$, respectively, for $p=7$ and $11$.}

\begin{table*}[t]
\vspace{-1ex}
\centering
\resizebox{\textwidth}{!}{
	\begin{tabular}{|c|c|c|c|c|c|c|c|c|c|c|c|c|c|c|c|c|c|c|}
		\hline
		\multirow{3}{*}{$p$}
		&\multicolumn{6}{c|}{$\mathbf{H}(3,p,L)$} 
		&\multicolumn{6}{c|}{$\mathbf{H}_{1}(4,p,L)$} 
		&\multicolumn{6}{c|}{{$\mathbf{H}_4(5,p,L)$}} \\

		\cline{2-19}
		
		&\multicolumn{3}{c|}{$A_1$}
		&\multicolumn{3}{c|}{$A_2$}
		&\multicolumn{3}{c|}{$A_1$}
		&\multicolumn{3}{c|}{$A_2$}
		&\multicolumn{3}{c|}{$A_1$}
		&\multicolumn{3}{c|}{$A_2$}\\
		
		\cline{2-19}
		
		&M1&M2&Ran&M1&M2&Ran&M1&M2&Ran&M1&M2&Ran&M1&M2&Ran&M1&M2&Ran \\
		\hline
		$5$&$0$&$0.6$&$1.33$&$0$&$0$&$0.67$&$2.80$&$4.80$&$5.06$&$0.80$&$0$&$2.66$&n/a&n/a&n/a&n/a&n/a&n/a \\
		\hline
		$7$&$0.43$&$0.86$&$1.48$&$0$&$0.57$&$0.64$&$4.71$&$3.43$&$8.12$&$2.85$&$1.46$&$4.57$
		&$20.98$&$17.70$&$29.83$&$19.21$&$9.35$&$23.26$ \\
		\hline
		$11$&$0.99$&$1.82$&$3.36$&$0$&$0.73$&$1.70$&$9.45$&$8.18$&$13.0$&$4.87$&$2.54$&$6.74$
		&$38.88$&$38.01$&$52.06$&$26.68$&$23.33$&$55.09$ \\
		\hline
	\end{tabular}
}
	\caption{\label{tab:tab1} Values of $A_1$ and $A_2$ obtained for nested AB-SC-LDPC matrices using Method 1 (M1), Method 2 (M2), and random generation (Ran), as $L\rightarrow \infty$. }
\end{table*}


By comparing the results obtained using Method 1 and 2 in Table \ref{tab:tab1} to the uncoupled matrices ($A^{(BC)}$ values), we note that spatial coupling is able to significantly reduce the number of 6-cycles, and hence dominant ABSs. Moreover, Method 1 results in a lower $A_m$ for the global SC-LDPC matrix $\mathbf{H}(3,p,L)$, but has a {larger} $A_m$ for the nested SC-LDPC matrices. {Using} Method 2, on the other hand, shows the opposite behavior - a {smaller} $A_m$ for the nested codes, but a {moderate} $A_m$ for the global code. From the Method 2 results shown in Table \ref{tab:tab1}, we also note that, for $p=5$ and $m=2$, the asymptotic average is zero in all optimized column weight-$3$ or column weight-$4$ nested codes. From Table \ref{tab:tab1}, we observe that, on average, the randomly generated nested SC-LDPC matrices possess a significantly larger value of $A_m$ (and hence they contain a larger number of dominant ABSs) when compared to nested AB-SC-LDPC matrices obtained via ALC based optimization, but still show a significant reduction when compared to AB-LDPC-BCs ($A^{(BC)}$ values). {Note that for $p=5$, the rate of the column weight-$5$ nested BC is zero, and hence the results for $\mathbf{H}_4(5,5,L)$ are excluded.}

Finally, we note that by applying the terminal lift described in Section \ref{sec:Opt_term}, even for small $J$, it is possible to significantly reduce, or even eliminate dominant ABSs. For example, with $p=5,7$, $m=1,2$, and $p=11,m=2$, and a lifting factor of only $J=5$, we are able to completely eliminate all 6-cycles. This means that, after a terminal lift of the AB-SC-LDPC nested codes obtained via Method 1 and 2 in Table \ref{tab:tab1} {the corresponding} $(3,3)$, $(4,4)$, $(5,4)$ and $(6,4)$ ABSs {as described in Remark \ref{rem:6cyc_33ABS}} are {removed}. {Note that, by removing 6-cycles in these AB-SC-LDPC codes, we are also able to significantly reduce the number of 8-cycles and any ABSs connected to them.}  

\vspace{-0.5cm}
\subsection{Simulated Code Performance}
\label{sec:results2}

We now consider the BP decoding performances of terminally lifted column weight-$3$, $4$, {and $5$} {QC}-SC-LDPC nested parity-check matrices. To demonstrate the behavior, we construct $\mathbf{H}(3,p,L=99,J=5)$, $\mathbf{H}_1(4,p,L=99,J=5)$, {and $\mathbf{H}_4(5,p,L=99,J=5)$} nested {QC}-SC-LDPC codes, with memory $m=2$ and {$p=7,11$}, as described in Section \ref{sec:results1}, and compare their performances to random (non-optimized before terminal lift) {QC}-SC-LDPC codes and terminally lifted nested {QC}-LDPC-BC parity-check matrices $\mathbf{H}(3,p,J'(m+1))$, $\mathbf{H}_1(4,p,J'(m+1))$, {and $\mathbf{H}_4(5,p,J'(m+1))$}. \hspace{-0.25cm} \footnote{{Note that $\mathbf{\hat{H}}(6,p)$ also consists of nested sub-matrices $\mathbf{H}_2(4,p)$, $\mathbf{H}_3(4,p)$, and $\mathbf{H}_5(5,p)$ with row groups $\{0,1,2,4\}$, $\{0,1,2,5\}$ and $\{0,1,2,3,5\}$, respectively, but the numerical results for $\mathbf{H}_{i'}(4,p,L)$, $i'\in \{2,3\}$, and $\mathbf{H}_5(5,p,L)$ are not shown as they are similar to the results for $\mathbf{H}_1(4,p,L)$, and $\mathbf{H}_4(5,p,L)$, respectively.}} 




Computer simulation results are obtained over a binary additive white Gaussian noise (AWGN) channel. A sliding window BP decoder is used for decoding nested SC codes \cite{IPSWVC12}, where the window attempts to decode one block of $p^2J$ target symbols in each decoding instant, then the decoder shifts one block right and one block down the parity-check matrix (\ref{eq:sc_mat}), thereby decoding the entire frame after $L$ shifts. The window length $W$ of the decoder was selected as $4\nu'$ symbols, and $50$ iterations were performed per window position. The terminally lifted {QC}-LDPC-BCs were decoded with a standard flooding BP decoder with a maximum of $50$ iterations and a syndrome check stopping rule. The block length of the short LDPC-BC, obtained using $J'=5$, is identical to the constraint length $\nu'=p^2J(m+1)$ of the SC code, whereas the long LDPC-BC, obtained using $J'=20$, has block length {equal} to the window length $4\nu'$. 


\vspace{-0cm}
\begin{figure*}[p]
\begin{subfigure}{0.5\textwidth}
  \centering
  \includegraphics[width=1\linewidth]{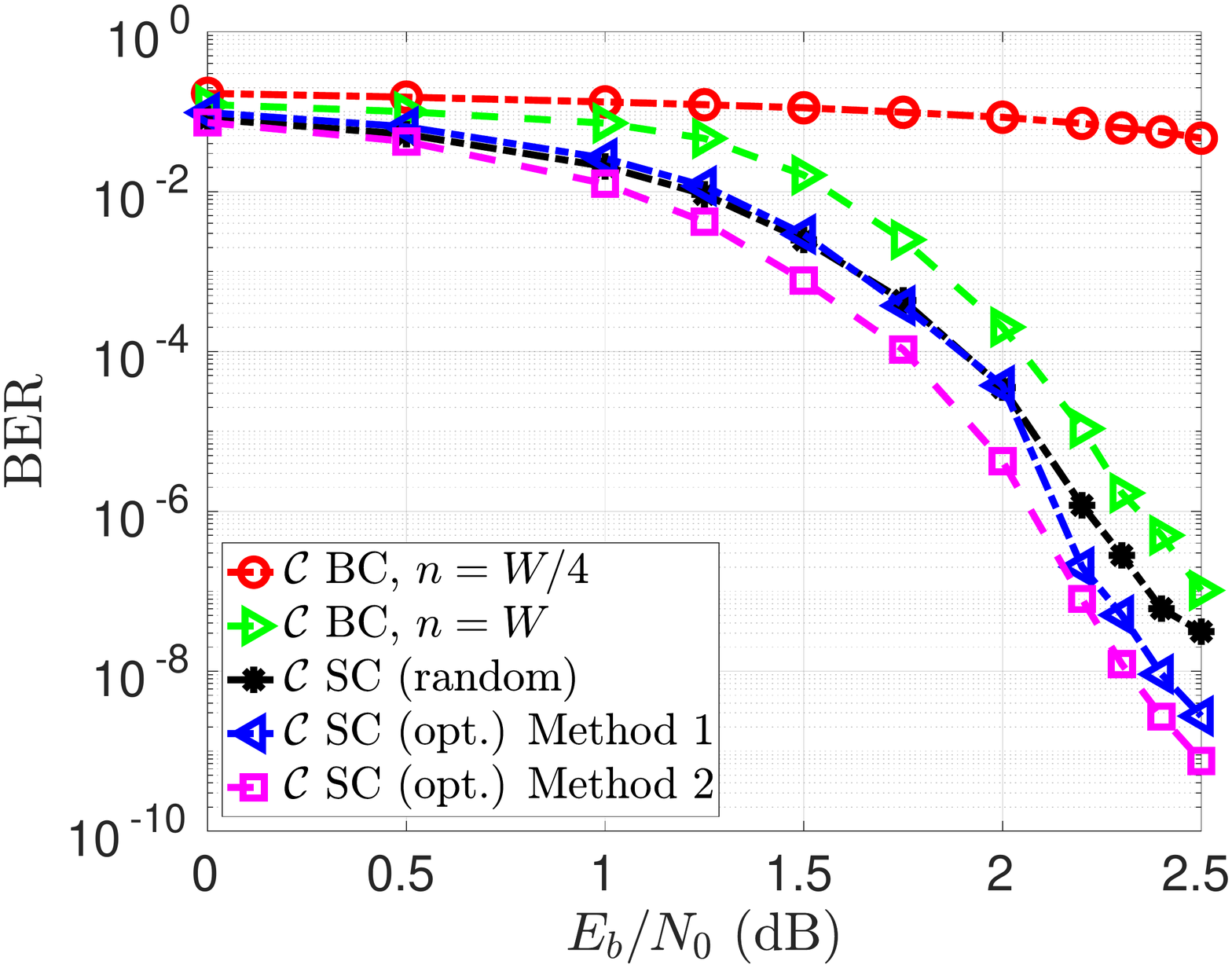}
  \caption{}
  \label{fig:res1}
\end{subfigure}
\begin{subfigure}{.5\textwidth}
  \centering
  \includegraphics[width=1\linewidth]{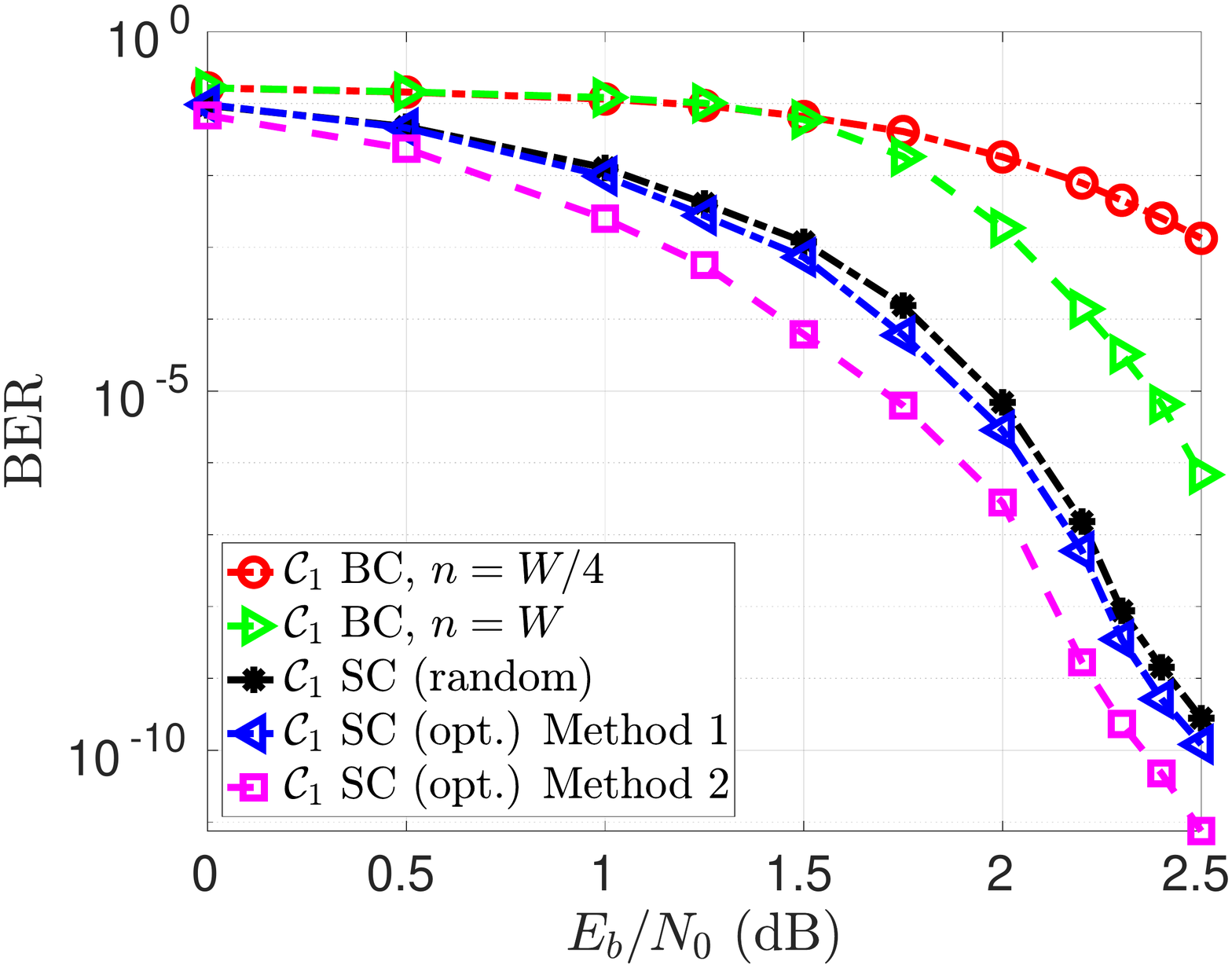}
  \caption{}
  \label{fig:res2}
\end{subfigure}
\begin{subfigure}{.5\textwidth}
  \centering
  \includegraphics[width=1\linewidth]{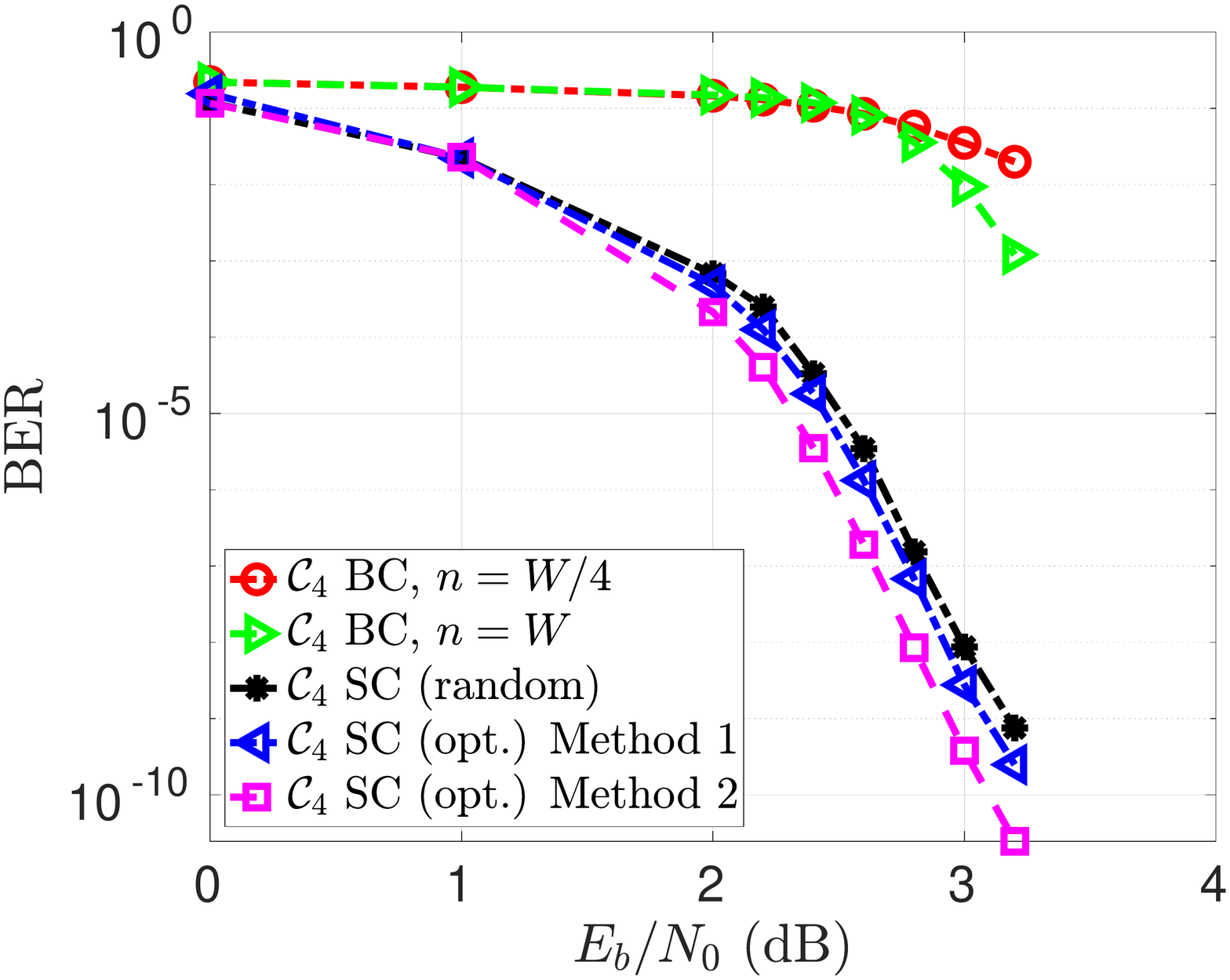}
  \caption{}
  \label{fig:res3}
\end{subfigure}
\begin{subfigure}{.5\textwidth}
  \centering
  \includegraphics[width=1\linewidth]{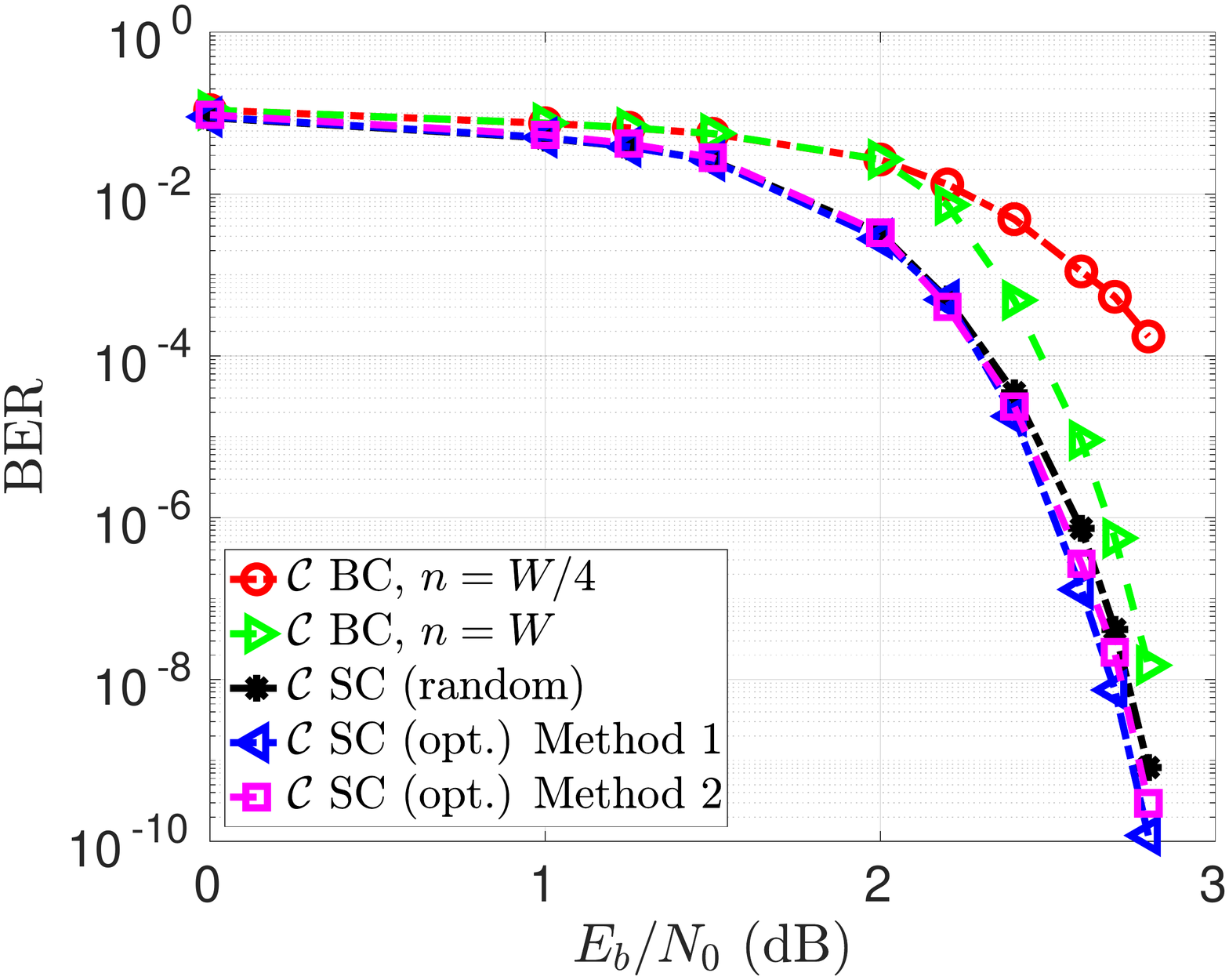}
  \caption{}
  \label{fig:res4}
\end{subfigure}
\begin{subfigure}{.5\textwidth}
  \centering
  \includegraphics[width=1\linewidth]{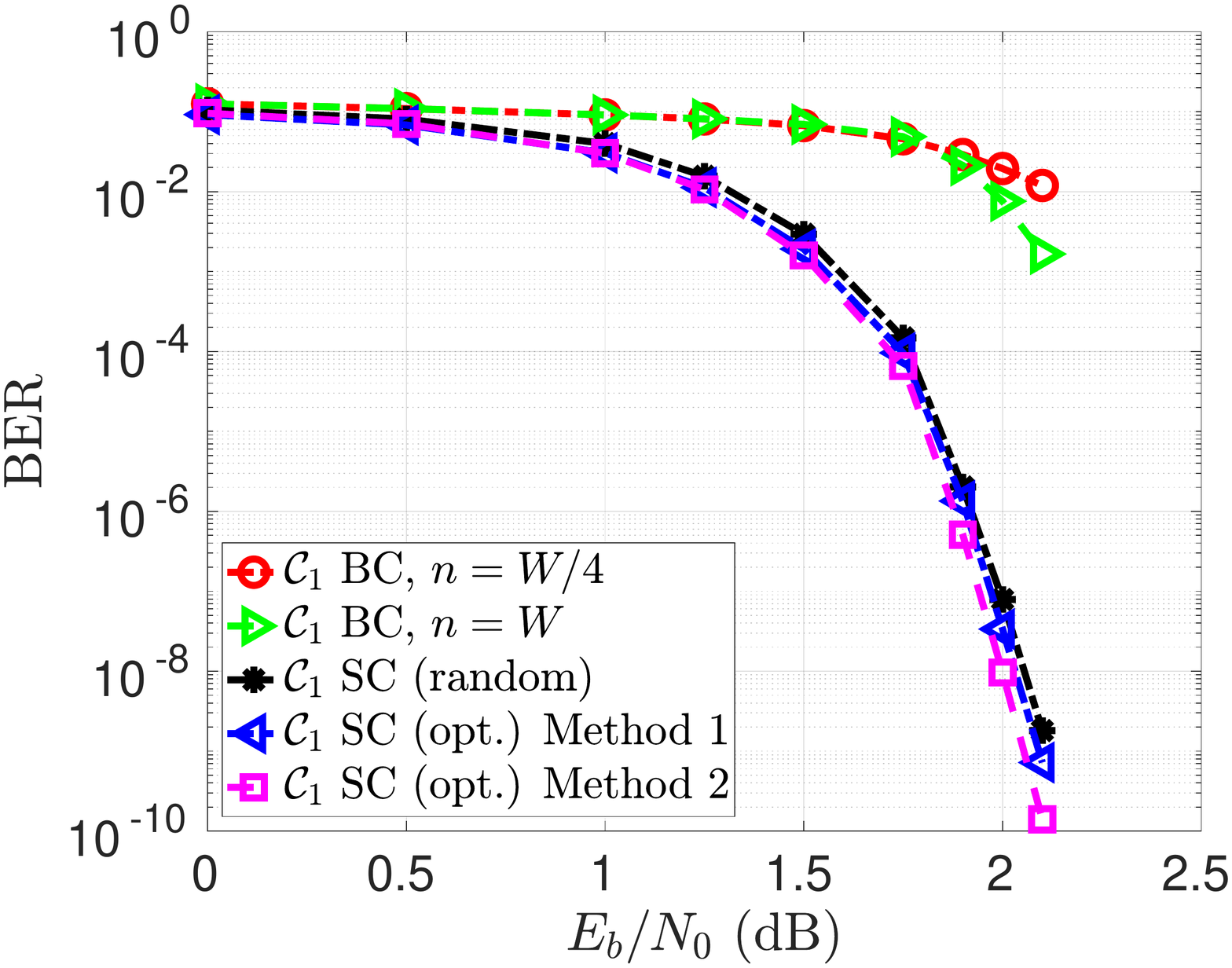}
  \caption{}
  \label{fig:res5}
\end{subfigure}
\begin{subfigure}{.5\textwidth}
  \centering
  \includegraphics[width=1\linewidth]{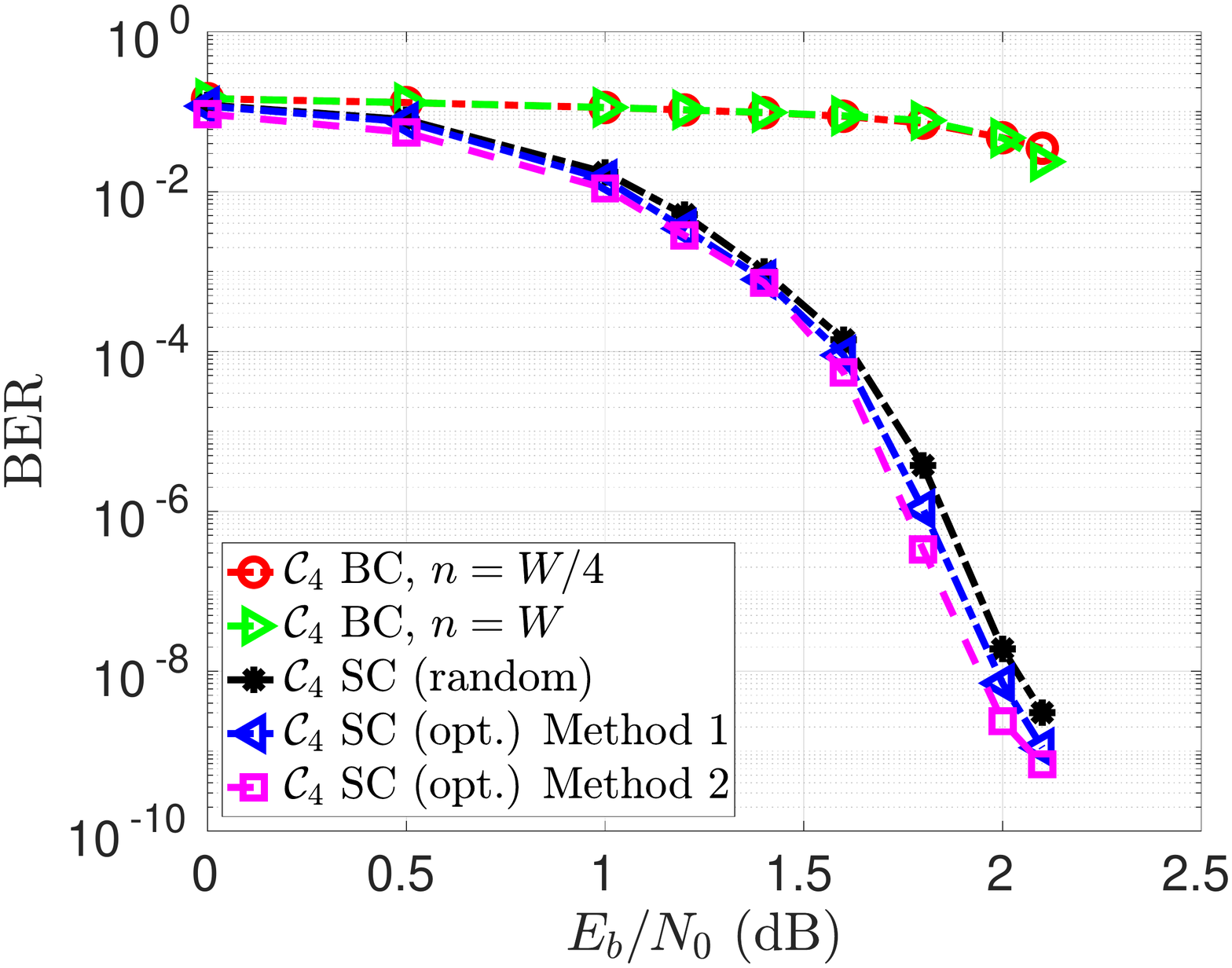}
  \caption{}
  \label{fig:res6}
\end{subfigure}
\vspace{-1ex}

\caption{{BER results obtained for nested QC-LDPC-BC and QC-SC-LDPC codes for $p=7$, and window size $W=2940$: (a) global codes, (b) column weight-$4$ nested codes, and (c) column weight-$5$ nested codes, and for $p=11$, and $W=7260$: (d) global codes, (e) column weight-$4$ nested codes, and (f) column weight-$5$ nested codes.} }
\vspace{-2.5ex}
\label{fig:curves}
\end{figure*}


The bit error rate (BER) results are shown in Fig. \ref{fig:curves}. As expected from the enumeration results in Section \ref{sec:results1}, the {performance} of the nested {QC}-SC-LDPC codes, {irrespective of the column weight or construction method}, {is} much better when compared to the performance of the corresponding nested regular {QC}-LDPC-BCs. We observe a significant gain in the waterfall for the optimized {QC}-SC-LDPC codes attributed to threshold saturation effect, and no indication of an error floor. The random SC-LDPC codes do appear to display the beginning of an error floor at $2.5$ dB {for $p=7$}. {When comparing the performance of the two optimization strategies, it is noticeable that both the optimized column weight-$4$ and $5$ nested QC-SC-LDPC codes perform better under Method 2 optimization when compared to Method 1. In case of the column weight-$3$ code, we see the same behavior for $p=7$, which can be attributed in part to the terminal lift and resulting code realization.} Note that the lower bound on the decoding error probability of message passing algorithms is known to decrease linearly with the multiplicity of harmful ABSs in SC codes \cite{HMCF20}. Hence, the optimized nested {QC}-SC-LDPC codes are expected to have better error-floor performance when constructed via Method 1 or 2 compared to randomly generated {QC}-SC-LDPC codes. 


\section{Conclusion}

In this paper, we have considered the construction of finite length nested AB-SC-LDPC codes from nested AB-LDPC-BCs. During code construction, we ensured that each nested SC sub-code and the global code have a small number of dominant ABSs in the Tanner graph when compared to the underlying LDPC-BCs. An ALC based optimization technique was utilized to optimize the design of nested AB-SC-LDPC codes. This optimization scheme allows the enumeration of dominant ABSs (by counting 6-cycles) in arbitrary column weight-$3$ sub-matrices of the nested parity-check matrices, facilitating a tractable nested code optimization. We demonstrate that, by using ALC, it is possible to minimize/eliminate dominant ABSs in certain nested AB-SC-LDPC codes, irrespective of the row and column weight of the overall parity-check matrix containing all the nested matrices. We also show that, for some specific codes, dominant absorbing sets in the Tanner graphs of all nested codes are completely removed using our proposed optimization strategy. Simulation results verify the improved nested code performance under ALC based optimization.

\vspace{-0.25cm}

\begin{appendices}

\section{Proof of Lemma \ref{lem:ALC}}
\label{app:range}

Recall the labeling of a 6-cycle shown in Fig. \ref{fig:6-cycle} in case of AB matrices with parameters $q_{i}$, $j_{i}$, and $k_{i}$, $i\in \{1,2,3\}$. The edges corresponding to $(r_{2},c_3)$ and $(r_{2},c_{2})$ both exist in row $s_2$, hence we get $s_2= q_2j_3+k_3 \text{ mod } p= q_2j_2+k_2 \text{ mod } p$. As a result,

\vspace{-0.5cm}
\begin{equation}
\label{eq:2_1}
q_2j_2+k_2 = q_2j_3+k_3 \ \text{ mod } p.
\end{equation} 

\vspace{-0cm}
\noindent Similarly, considering $(r_2,c_3)$ and $(r_3,c_1)$ it is noted that $s_3 = q_3j_1+k_1 \text{ mod } p = q_3j_3+k_3 \text{ mod } p$, which gives

\vspace{-0.25cm}
\begin{equation}
\label{eq:3-1}
q_3j_1+k_2= q_3j_3+k_3 \ \text{ mod } p,
\end{equation} 

\vspace{-0cm}
\noindent where $j_1<j_2$. Note that (\ref{eq:2_1}) and (\ref{eq:3-1}) can be rearranged as 

\vspace{-0.25cm}
\begin{equation}
k_3-k_2= q_2j_2-q_2j_3 \text{ mod } p, \label{eq:6b}
\end{equation} 

\vspace{-0.25cm}
\noindent and 

\vspace{-0.25cm}
\begin{equation}
k_3-k_2= q_3j_1-q_3j_3 \text{ mod } p, \label{eq:6b-1}
\end{equation} 

\vspace{-0.2cm}
\noindent respectively. Equations (\ref{eq:6b}) and (\ref{eq:6b-1}) can also be written as

\vspace{-0.25cm}
\begin{equation}
k_3-k_2= q_2j_2-q_2j_3 -\lambda_1 p, \label{eq:6c}
\end{equation} 

\vspace{-0.25cm}
\noindent and 

\vspace{-0.25cm}
\begin{equation}
k_3-k_2= q_3j_1-q_3j_3 -\lambda_2 p, \label{eq:6c-1}
\end{equation} 

\noindent respectively, where $\lambda_1,\lambda_2\in \{1-p,2-p,\ldots,p-1\}$. By substituting (\ref{eq:6c}) in (\ref{eq:6c-1}), and after rearranging we get

\vspace{-0.1cm}
\begin{equation}
j_3=\frac{q_2j_2-q_3j_1-\lambda p}{q_2-q_3},
\label{eq:10}
\end{equation} 

\noindent where $j_2>j_1$, $q_2,q_3\in \{1,2,\ldots,\gamma-1\}$ and $\lambda=\lambda_1-\lambda_2$. In general, $\alpha \leq j_3 \leq \beta-1$, where $0\leq\alpha\leq p-1$, $1\leq\beta\leq p$ and $\alpha<\beta$. Using this inequality and from (\ref{eq:10}) we have

\vspace{-0cm}
\begin{equation}
\alpha\leq \frac{q_2j_2-q_3j_1-\lambda p}{q_2-q_3} \leq \beta-1. \label{eq:10-1}
\end{equation}  

\vspace{-0cm}
\noindent From (\ref{eq:10-1}) we obtain the range of values of column group $j_3$ which contains column $c_3$ of the AB parity-check matrix. Multiplying through (\ref{eq:10-1}) by $p$ and using the relations $j_1p=c_1-k_1$ and $j_2p=c_2-k_2$, $c_2>c_1$, yields 

\vspace{-0.25cm}
\begin{align}
\label{eq:13b}
\begin{split}
 \alpha p \leq \frac{q_2c_2-q_3c_1+q_3k_1-q_2k_2-\lambda p^2}{q_2-q_3} \leq \beta p - p.
\end{split}
\end{align} 

\vspace{-0cm}
\noindent Now, considering that both $c_1$ and $c_2$ are associated with a row group of $\mathbf{I}$ matrices only, we obtain $k_1=k_2$. Then (\ref{eq:13b}) simplifies as

\vspace{-0.5cm}
\begin{align}
\label{eq:15}
\begin{split}
\alpha p \leq \frac{q_2c_2-q_3c_1+(q_3-q_2)k_1-\lambda p^2}{q_2-q_3} \leq \beta p - p.
\end{split}
\end{align} 

\vspace{-0cm}
\noindent Simplifying (\ref{eq:15}) further yields

\vspace{-0.25cm}
\begin{align}
\label{eq:16}
\begin{split}
\alpha p \leq \frac{q_2c_2-q_3c_1-\lambda p^2}{q_2-q_3}-k_1 \leq \beta p - p \\
\Rightarrow \alpha p+k_1 \leq \frac{q_2c_2-q_3c_1-\lambda p^2}{q_2-q_3} \leq \beta p - p+k_1. \\
\end{split}
\end{align}

\vspace{-0cm}
\noindent Since $0\leq k_{1}<p$, (\ref{eq:16}) is rewritten as 

\vspace{-0.25cm}
\begin{equation}
\label{eq:19}
\begin{split}
\alpha p \leq \frac{q_2c_2-q_3c_1-\lambda p^2}{q_2-q_3} < \beta p. 
\end{split}
\end{equation}

\vspace{-0cm}
\noindent By rearranging (\ref{eq:19}), the range of $c_3$ is expressed in terms of $c_1$ and $c_2$ as
	\begin{align}
	\label{eq:ineq2b}
	\begin{split}	
		\left(1-\frac{q_3}{q_2}\right)\alpha p+ \frac{\lambda p^2}{q_2}\leq c_2-\frac{q_3}{q_2}c_1<\left(1-\frac{q_3}{q_2}\right)\beta p+ \frac{\lambda p^2}{q_2}, 		
	\end{split}
	\end{align}

\vspace{-0.15cm}
\noindent where $c_2>c_1$, $\lambda\in\{2-2p,3-2p,\ldots,2p-2\}$, $q_2,q_3\in\{1,\ldots,\gamma-1\}$ and $q_2\neq q_3$. \hfill\(\Box\)

\section{Proof of Proposition \ref{prop:alc}}
\label{app:ALC}

\begin{figure}[h!]
\begin{subfigure}{.5\textwidth}
  \centering
  \centerline{\resizebox{3.15in}{2.3in}{\includegraphics{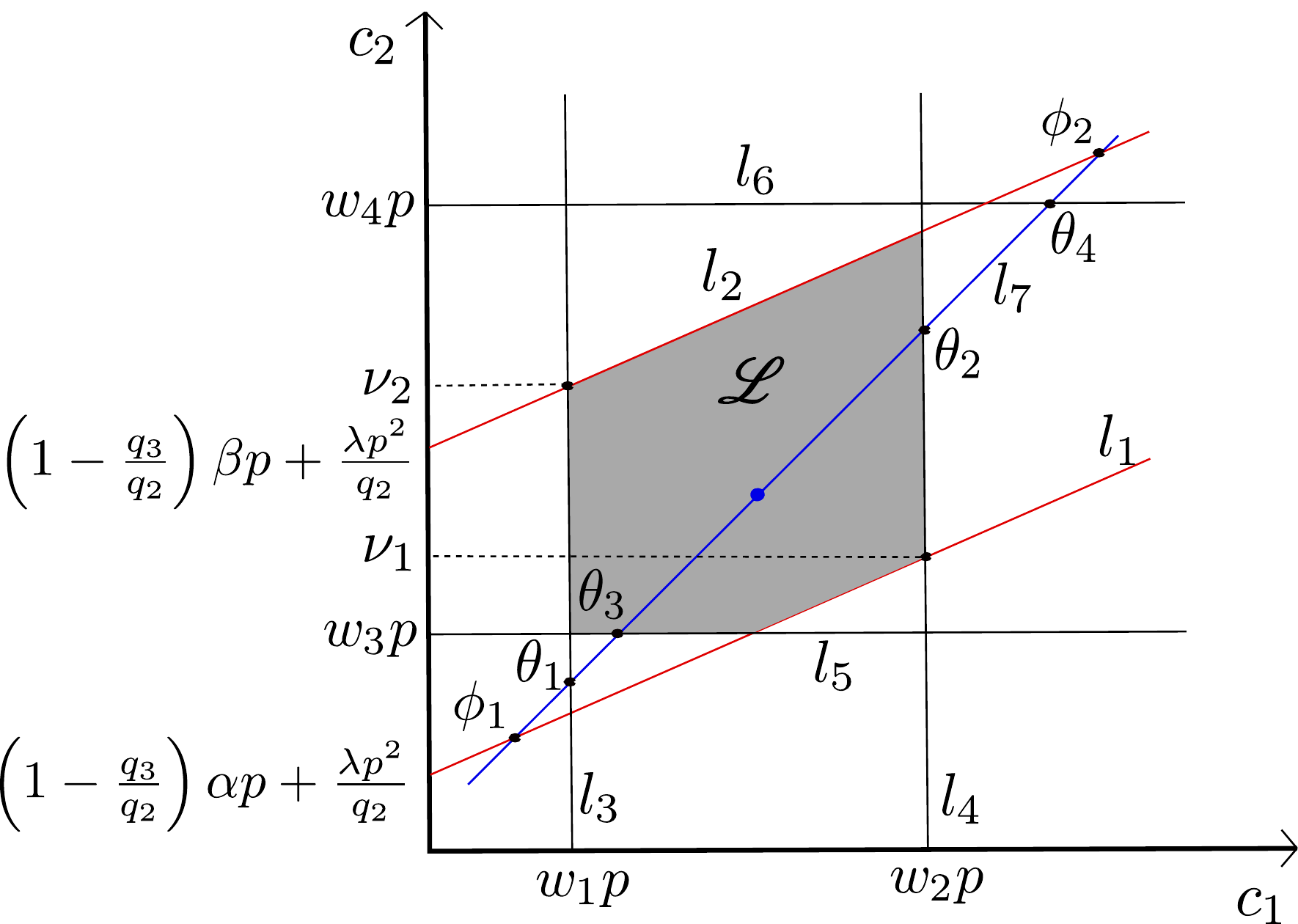}}}
  \caption{}
  \label{fig:N0R}
\end{subfigure}
\begin{subfigure}{.5\textwidth}
  \centering
  \centerline{\resizebox{2.52in}{2.3in}{\includegraphics{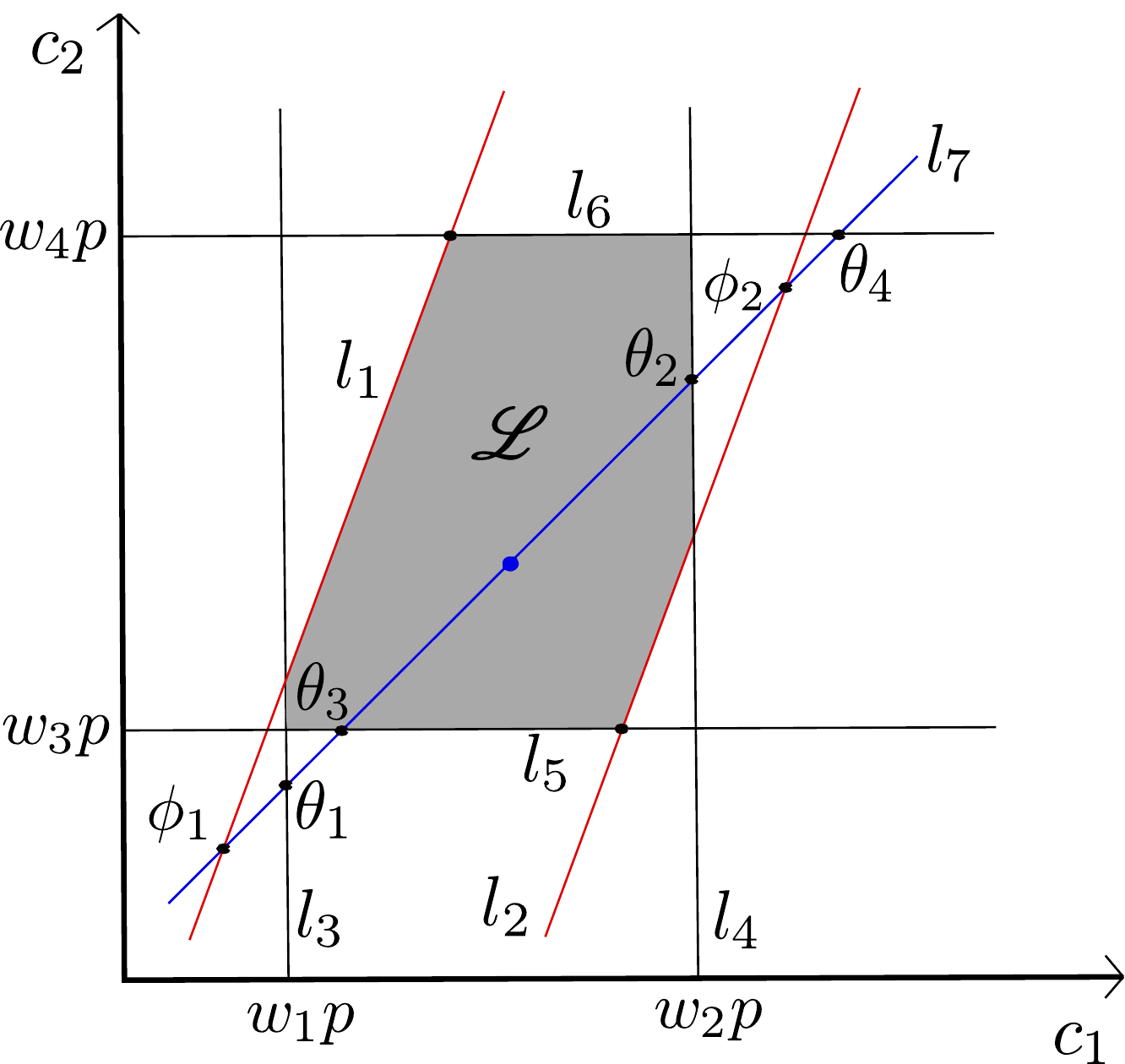}}}
  \caption{}
  \label{fig:N1R}
\end{subfigure}
\vspace{-0.25cm}
\caption{Illustration of the $(c_1,c_2)$ plane for the case (a) $q_3<q_2$ and (b) $q_3>q_2$. The blue point on $l_7$ represents a 6-cycle.}
\vspace{-0.5cm}
\label{fig:LC_figs}
\end{figure}

From the lower bound of (\ref{eq:ineq2b}), we obtain a line $c_2=\frac{q_3}{q_2}c_1+\left(1-\frac{q_3}{q_2}\right)\alpha p+ \frac{\lambda p^2}{q_2}$ on the $(c_1,c_2)$ plane, where $\frac{q_3}{q_2}$ is the gradient of the line and $\left(1-\frac{q_3}{q_2}\right)\alpha p+ \frac{\lambda p^2}{q_2}$ is its vertical intercept. Let this line be labeled as $l_1$. Similarly, from the upper bound of (\ref{eq:ineq2b}) we obtain another line $c_2=\frac{q_3}{q_2}c_1+\left(1-\frac{q_3}{q_2}\right)\beta p+ \frac{\lambda p^2}{q_2}$, and let this line be labeled as $l_2$. Now, let $l_3,l_4,l_5,l_6$, and $l_7$, respectively, be the labels of the lines $c_1=w_1p$, $c_1=w_2p$, $c_2=w_3p$, $c_2=w_4p$ and $c_2-c_1=np$, on the $(c_1,c_2)$ plane. These lines are shown in Figs. ~\ref{fig:LC_figs}(a) and \ref{fig:LC_figs}(b).\footnote{Note that Figs. \ref{fig:LC_figs}(a) and \ref{fig:LC_figs}(b) reveal the difference in steepness of lines $l_1$ and $l_2$ as their gradient ($q_3/q_2$) changes from less than $1$ in Fig. \ref{fig:LC_figs}(a), to more than $1$ in Fig. \ref{fig:LC_figs}(b).} Recall from Section \ref{sec:LC} that a $(c_1,c_2)$ integer pair on $l_7$ in the grey region, $\mathscr{L}$, represents a 6-cycle in the corresponding matrix region $\mathcal{R}$. Hence, the total number of 6-cycles in $\mathcal{R}$ is equal to the length of line $l_7$ within $\mathscr{L}$.

%

\vspace{-0cm}


\vspace{-0cm}
Let $\phi_{1}$, $\phi_{2}$, $\theta_1$, $\theta_2$, $\theta_3$, and $\theta_4$, be the points of intersection of line $l_7$ with lines $l_1$, $l_2$, $l_3$, $l_4$, $l_5$, and $l_6$, respectively. Note that the points $\phi_1,\theta_1,\theta_3$ and $\phi_2,\theta_2,\theta_4$ are close to each other on the $(c_1,c_2)$ plane, and hence we call them ``neighbors''. Any two points, selected from each neighbor, producing the shortest length of $l_7$ within the rectangular boundary imposed by $l_{3},l_{4},l_{5}$, and $l_{6}$, are denoted as $\sigma_{1}$ and $\sigma_{2}$. That is, $\sigma_{1x}=\max(\phi_{1x},\theta_{1x},\theta_{3x})$, $\sigma_{1y}=\max(\phi_{1y},\theta_{1y},\theta_{3y})$, $\sigma_{2x}=\min(\phi_{2x},\theta_{2x},\theta_{4x})$, $\sigma_{2y}=\min(\phi_{2y},\theta_{2y},\theta_{4y})$, where the $x$ and $y$ subscripts indicate the $x$ and $y$ coordinates, respectively, on the $(c_1,c_2)$ plane. Hence, in Figs. \ref{fig:LC_figs}(a) and (b), $\sigma_1=\theta_3$ and $\sigma_2=\theta_2$. From the principles of Cartesian geometry, the number of integer points lying on straight line connecting $\sigma_{1}$ and $\sigma_{2}$ is given as $\sqrt{\frac{(\sigma_{2x}-\sigma_{1x})^{2}+(\sigma_{2y}-\sigma_{1y})^{2}}{2}}$. Let the y-coordinates of the points of intersection between lines $l_4$ and $l_1$, and lines $l_3$ and $l_2$, be denoted as $\nu_1$ and $\nu_2$, respectively. Then, the total number of 6-cycles in $\mathcal{R}$ is expressed as 

\vspace{-0.25cm}
\begin{equation}
\mathcal{N}_{\mathcal{R}}=
\begin{cases}
\sqrt{\frac{(\sigma_{2x}-\sigma_{1x})^{2}+(\sigma_{2y}-\sigma_{1y})^{2}}{2}}, & \text{if (\ref{eq:conditions}) hold}, \\
0, & \text{otherwise}.
\end{cases}
\end{equation}

\vspace{-0cm}
\noindent The boundary conditions that ensure line $l_7$ will intersect the shaded region $\mathscr{L}$ in Fig. \ref{fig:LC_figs} are \\

\vspace{-0.5cm}
\begin{align}
	\begin{split}
		\theta_{1y}<\nu_2, \theta_{2y}>\nu_1 \text{ if } q_3<q_2, \text{ or } \\  
w_3 p<\phi_{2y}, w_4 p>\phi_{1y} \text{ if } q_3>q_2, \text{ and } \\   
\theta_{1x}\leq\{\sigma_{1x},\sigma_{2x}\}\leq\theta_{2x},\theta_{1y}\leq\{\sigma_{1y},\sigma_{2y}\}\leq\theta_{2y},
	\end{split}
	\label{eq:conditions}
\end{align}

\vspace{-0.1cm}
\noindent where $\nu_1=\alpha p +(w_2-\alpha)\frac{q_3}{q_2}p+\lambda \frac{p^2}{q_2}$, $\nu_2=\beta p +(w_1-\beta)\frac{q_3}{q_2}p+\lambda \frac{p^2}{q_2}$, $\phi_{1y}=\frac{q_2}{q_2-q_3}\left[\alpha p- \frac{q_3}{q_2}p(n+\alpha)+\frac{\lambda p^2}{q_2}\right]$, and $\phi_{2y}=\frac{q_2}{q_2-q_3}\left[\beta p- \frac{q_3}{q_2}p(n+\beta)+\frac{\lambda p^2}{q_2}\right]$. \hfill\(\Box\)

\section{Proof of Proposition \ref{prop:Nubound}}
\label{app:Nubound}

	{A 6-cycle exists in two possible region configurations within $m+1$ contiguous column blocks of a column weight-$3$ AB-SC-LDPC parity-check matrix. The first configuration is a single column block $\mathbf{H}'=[\mathbf{H}_0^\top \text{ } \mathbf{H}_1^\top, \cdots, \text{ } \mathbf{H}_m^\top]^\top$, and the other one is an  ``L'' shaped matrix region configuration
	
\vspace{-0.15cm} 
\begin{align}
\mathbf{H}^{\dprime}&=\begin{bmatrix}
\mathbf{H}_0 & \\
\mathbf{H}_1 & \mathbf{H}_0 \\
\vdots & \vdots & \ddots \\
\mathbf{H}_m & \mathbf{H}_{m-1} & \cdots & \mathbf{H}_0\\
\end{bmatrix}, 
\text { or } \\
\mathbf{H}^{\tprime}&=\begin{bmatrix}
\mathbf{H}_m & \mathbf{H}_{m-1} & \cdots & \mathbf{H}_0\\
 & \mathbf{H}_m & \cdots & \mathbf{H}_1 \\
 & & \ddots  & \vdots \\
 & & & \mathbf{H}_m \\
\end{bmatrix}.
\end{align}

Fix a single row group of $\mathbf{I}$ (identity) matrices within $\mathbf{H}'$, and randomly select a pair of row group indices from the remaining $2(m+1)$ non-$\mathbf{I}$ matrix row group indices in $\mathbf{H}'$ to create a triple (representing the row group indices of a column weight-$3$ sub-matrix in $\mathbf{H}'$). There are $\binom{2(m+1)}{2}$ possible ways of generating these triples, with a common row group index of $\mathbf{I}$ matrices. However, $2\binom{m+1}{2}$ of these triples contain a pair of non-$\mathbf{I}$ matrix row group indices $(q,q')$ such that $q =q' \text{ mod } 3$, and row groups with these triples do not contain any 6-cycles. Thus, for a given index of an $\mathbf{I}$-matrix row group in $\mathbf{H}'$, we obtain $\kappa= \binom{2(m+1)}{2}-2\binom{m+1}{2}$ distinct column weight-$3$ matrix regions containing 6-cycles. Since there are $m+1$ row group of $\mathbf{I}$ matrices in $\mathbf{H}'$, the total number of distinct column weight-$3$ matrix regions in $\mathbf{H}'$ (containing 6-cycles) is $(m+1)\kappa$.}

{Now, note that the combination of the two ``L'' shaped matrix regions have in total $2m+1$ row groups containing $\mathbf{I}$ matrices only, {and hence there are $(2m+1) \kappa$ column weight-$3$ matrix regions (containing 6-cycles) which span one or more column blocks of $\mathbf{H}^{\dprime}$ or $\mathbf{H}^{\tprime}$. Since at least one of these regions is already in $\mathbf{H}'$}, the number of distinct column weight-$3$ matrix regions (containing 6-cycles) spanning more than one column block of $\mathbf{H}^{\dprime}$ or $\mathbf{H}^{\tprime}$ is upper-bounded by $(2m+1) \kappa$. Therefore, we obtain $N<(m+1)\kappa+(2m+1) \kappa=(3m+2)\kappa$.} \hfill\(\Box\)


\end{appendices}

\bibliographystyle{IEEEtran}
\vspace{-0.25cm}
{\linespread{0.9}\selectfont\bibliography{references}}

\end{document}